\documentclass[aps,prl,twocolumn,superscriptaddress,longbibliography]{revtex4-1}

\usepackage{graphicx}
\usepackage{amsmath}
\usepackage{color}
\usepackage[colorlinks=true,linkcolor=blue,urlcolor=blue,citecolor=blue]{hyperref}
\usepackage{hyperref}
\usepackage{braket}
\usepackage{amssymb}
\usepackage{tikz}

\begin{document}
%\preprint{}

\title{Observation of Slow Dynamics near the Many-Body Localization Transition in One-Dimensional Quasiperiodic Systems}

\author{Henrik P. L\"uschen}
\affiliation{Fakult\"at f\"ur Physik, Ludwig-Maximilians-Universit\"at M\"unchen, Schellingstr.\ 4, 80799 Munich, Germany}
\affiliation{Max-Planck-Institut f\"ur Quantenoptik, Hans-Kopfermann-Str.\ 1, 85748 Garching, Germany}

\author{Pranjal Bordia}
\affiliation{Fakult\"at f\"ur Physik, Ludwig-Maximilians-Universit\"at M\"unchen, Schellingstr.\ 4, 80799 Munich, Germany}
\affiliation{Max-Planck-Institut f\"ur Quantenoptik, Hans-Kopfermann-Str.\ 1, 85748 Garching, Germany}

\author{Sebastian Scherg}
\affiliation{Fakult\"at f\"ur Physik, Ludwig-Maximilians-Universit\"at M\"unchen, Schellingstr.\ 4, 80799 Munich, Germany}
\affiliation{Max-Planck-Institut f\"ur Quantenoptik, Hans-Kopfermann-Str.\ 1, 85748 Garching, Germany}

\author{Fabien Alet}
\affiliation{Laboratoire de Physique Th{\'e}orique, IRSAMC, Universit{\'e} de Toulouse, CNRS, 31062 Toulouse, France}

\author{Ehud Altman}
\affiliation{Department of Condensed Matter Physics, Weizmann Institute of Science, Rehovot 7610001, Israel}
\affiliation{Department of Physics, University of California, Berkeley, CA 94720}

\author{Ulrich Schneider}
\affiliation{Fakult\"at f\"ur Physik, Ludwig-Maximilians-Universit\"at M\"unchen, Schellingstr.\ 4, 80799 Munich, Germany}
\affiliation{Max-Planck-Institut f\"ur Quantenoptik, Hans-Kopfermann-Str.\ 1, 85748 Garching, Germany}
\affiliation{Cavendish Laboratory, University of Cambridge, Cambridge CB3 0HE, UK}

\author{Immanuel Bloch}
\affiliation{Fakult\"at f\"ur Physik, Ludwig-Maximilians-Universit\"at M\"unchen, Schellingstr.\ 4, 80799 Munich, Germany}
\affiliation{Max-Planck-Institut f\"ur Quantenoptik, Hans-Kopfermann-Str.\ 1, 85748 Garching, Germany}

\date{\today}

\begin{abstract}

In the presence of sufficiently strong disorder or quasiperiodic fields, an interacting many-body system can fail to thermalize and become many-body localized. The associated transition is of particular interest, since it occurs not only in the ground state but over an extended range of energy densities. So far, theoretical studies of the transition have focused mainly on the case of true-random disorder.
In this work, we experimentally and numerically investigate the regime close to the many-body localization transition in quasiperiodic systems. We find slow relaxation of the density imbalance close to the transition, strikingly similar to the behavior near the transition in true-random systems. This dynamics is found to continuously slow down upon approaching the transition and allows for an estimate of the transition point. We discuss possible microscopic origins of these slow dynamics.
\end{abstract}

\pacs{}
%\keywords{}

\maketitle


\paragraph{\bf{Introduction.---}}

An isolated quantum system of interacting particles can be nonergodic and fail to thermalize in the presence of sufficiently strong disorder~\cite{Basko06,Polyakov05,Imbrie16,Oganesyan05,Pal10,Vosk13,Serbyn14,Shlyapnikov15,Nandkishore15,Altman15,Bahri15,Ovadia15,Schreiber15,Kondov15,Smith15,Choi16} or quasiperiodic fields~\cite{Roati08,Iyer13,Schreiber15}. This phenomenon -- called many-body localization (MBL) -- presents a generic alternative to thermalization~\cite{Deutsch91,Sred94,Rigol08} and has attracted an immense amount of interest in recent years; see, e.g., Refs.~\cite{Nandkishore15,Altman15} for reviews. More recently, theoretical studies started to address the phase transition from the thermalizing to the MBL phase itself (reviewed in Refs.~\cite{Vasseur16,Luitz16_3,Agarwal16}). This transition is of particular interest, since, in contrast to conventional quantum phase transitions~\cite{Sachdev_book} the MBL transition happens over a wide range of energy densities. Furthermore, a good understanding of the transition may give new insight into thermalization in closed quantum systems~\cite{Luitz16_2}.

So far, theoretical studies of the transition have focused on spin models with true-random disorder where the nature of the transition is still under discussion~\cite{Khemani16}.  Renormalization group schemes~\cite{Vosk15,Potter15} have predicted a Griffiths regime~\cite{Griffiths69} on the thermal side of the transition. In this regime, the dynamics is dominated by rare, locally critical or insulating inclusions in the thermalizing bulk, resulting in subdiffusive transport and power-law relaxation of global density patterns. 
Indeed, exact diagonalization (ED) studies of small systems have found slow power-law relaxation processes close to the MBL transition~\cite{Agarwal15,Lev15,TorresHerrera15,Luitz16,Znidaric16}, but with scaling behaviors in violation of the Harris-Chayes criterion~\cite{Harris74,Chayes86,Chandran15}. This is potentially due to finite size limitations preventing access to the scaling regime, suggesting that current numerics cannot accurately capture the properties of the true-random MBL transition~\cite{Khemani16}.
Recently, however, it has been pointed out that finite size limitations might be less severe in quasiperiodic systems~\cite{Khemani17}, as rare regions should \emph{a priori} be absent in a deterministic potential~\cite{Gopalakrishnan16}.

In this work, we experimentally and numerically investigate the MBL transition in a one-dimensional Fermi-Hubbard model with a quasiperiodic on-site potential. We find a slow relaxation dynamics of the density imbalance~\cite{Schreiber15} on the experimentally accessible time scales. These dynamics continuously slow down upon approaching the transition before stopping in the MBL phase, a behavior which is strongly reminiscent of a recent numerical study on true-random systems~\cite{Luitz16}. As an important result of the analysis of the dynamics, we are able to give an improved estimate of the critical point compared to previous values~\cite{Schreiber15}. Finally, we discuss possible microscopic explanations for the observed slow dynamics, including both rare regions in the initial state~\cite{Luitz16} and atypical transition rates between single-particle states~\cite{BarLev17}.
\\

\paragraph{\bf{Experiment.---}}

Our experimental setup effectively implements the interacting Aubry-Andr{\'e} model~\cite{Aubry80,Iyer13}, which describes spinful fermions on a lattice with nearest-neighbor tunneling of amplitude $J \approx h \times 500\,$Hz and on-site interactions of strength $U$. The fermions are subjected to a quasiperiodic correlated disorder potential of the form $\Delta \cos(2 \pi \beta i + \phi)$, where $\Delta$ and $\phi$ denote the strength and relative phase of the potential, $i$ numbers the lattice sites and the irrational $\beta$ gives the disorder periodicity (see Ref.~\cite{SOMs} for details). 
This model has a localization transition at $\Delta_c^{U=0} = 2 \, J$ in the absence of interactions~\cite{Aubry80}, and was shown numerically and experimentally to exhibit MBL above a critical disorder strength~\cite{Schreiber15}.

We prepare a high energy initial charge-density wave (CDW) state, where up and down spin atoms are randomly distributed on even lattice sites, while odd lattice sites are empty. During the preparation, doubly occupied lattice sites are suppressed by strong repulsive interactions. The CDW in the central tube is approximately 200 sites long and contains about 80 atoms. In contrast to previous experiments~\cite{Schreiber15}, in this work we only mildly confine the atom cloud during the ensuing time evolution in order to reduce the effects of the overall harmonic trapping potential. After a variable evolution time, we extract the imbalance $\mathcal{I} = (N_\mathrm{e} -N_\mathrm{o}) / (N_\mathrm{e} + N_\mathrm{o})$ between the populations of even ($N_\mathrm{e}$) and odd ($N_\mathrm{o}$) sites using a band mapping technique~\cite{Trotzky12}. The imbalance has an initial value close to one and, in a thermalizing system, will ultimately relax to zero. In contrast, a finite imbalance indicates a memory of the initial state and signals that the system has not fully thermalized yet. 
Since the imbalance is a local probe and does not require global mass transport to relax, it exhibits a short intrinsic relaxation timescale of $\mathcal{O}(\tau)$ in the nondisordered case, where $\tau = \hbar / J$ is the tunneling time. This allows for an experimental observation of slow, disorder induced dynamics. Global observables, on the other hand, are expected to show hydrodynamic tails in the ergodic phase~\cite{Rosch14}, which would mask the slow relaxation processes.
For details of the setup and the experimental sequence, see Refs.~\cite{Schreiber15,SOMs}.
\\

\begin{figure}
	\centering
	\includegraphics[width=84mm]{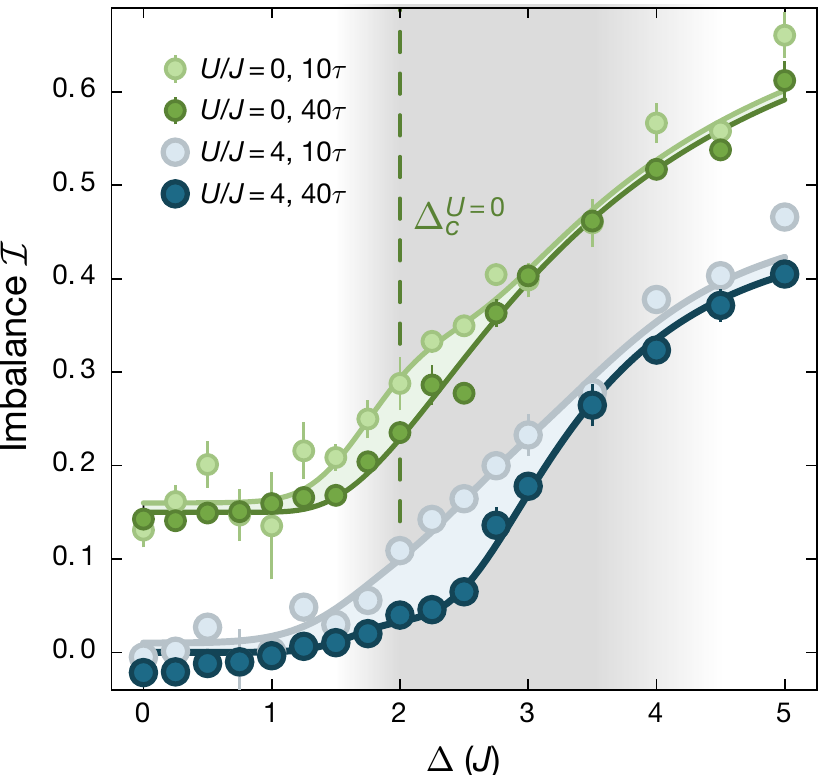}
	\caption{\textbf{Imbalance at finite times:} Measurements of the imbalance $\mathcal{I}$ after $10 \, \tau$ (light points) and $40 \, \tau$ (dark points) for the noninteracting system and at $U = 4 \, J$. The noninteracting data is vertically offset by $0.15$ for clarity. The data represents averages over 12 disorder phases $\phi$ with error bars indicating the uncertainty of the mean. Solid lines are guides to the eye.
		In the interacting system, we observe a regime (gray shaded), where the imbalance after $40 \, \tau$ is significantly lower than after $10 \, \tau$, indicating a dynamical evolution of the system. A similar, but much less pronounced, feature is also present in the noninteracting case.}
	\label{imb_fig}
\end{figure}

\paragraph{\bf{Finite-Time Imbalance.---}}
Fig.~\ref{imb_fig} shows measurements of the imbalance at various disorder strengths $\Delta$ for both the noninteracting case and at an interaction strength of $U=4 \, J$. The measurements were taken after $10 \, \tau$ (called short), which is nonetheless long enough for a clean system to relax, and after $40 \, \tau$ (called long). In this work, we generally refrain from accessing imbalances at times longer than $40 \, \tau$, since then background decays, which limit the lifetime of the imbalance to $\mathcal{O}(10^3 \, \tau)$, become increasingly relevant~\cite{SOMs,Bordia16,Lueschen16}. 

From the interacting data we can distinguish three different regimes, as indicated by the gray background shading. In the regimes of weak ($\Delta \lesssim 1.5 \, J$) and strong ($\Delta \gtrsim 4 \, J$) disorder, the imbalances measured after short and long times agree up to the effect of background decays~\cite{SOMs,Bordia16,Lueschen16}. The weak disorder regime is thermal, with the imbalance quickly relaxing to zero. The strong disorder regime shows many-body localization indicated by a rapid approach of the imbalance to a finite stationary value.
In the gray shaded regime of intermediate disorder strength ($1.5 \, J  \lesssim \Delta \lesssim 4 \, J$), we observe a significant difference between the interacting short and long term imbalance, indicating the presence of relaxation dynamics on a slow timescale. A similar trend, but much less pronounced, also exists in the noninteracting case in the vicinity of $\Delta_c^{U=0}$.
The fact that this regime extends to larger disorder strengths in the interacting case compared to the noninteracting case demonstrates that interactions give rise to an additional relaxation (thermalization) process. This additional process acts in addition to the critical slowing down present close to the noninteracting localization transition and hence shifts the MBL transition point to larger disorder strengths.

In the following, we present a detailed characterization of the slow dynamics in the interacting system. The equivalent analysis of the noninteracting system can be found in the supplemental material~\cite{SOMs}.
\\

\begin{figure}
	\centering
	\includegraphics[width=84mm]{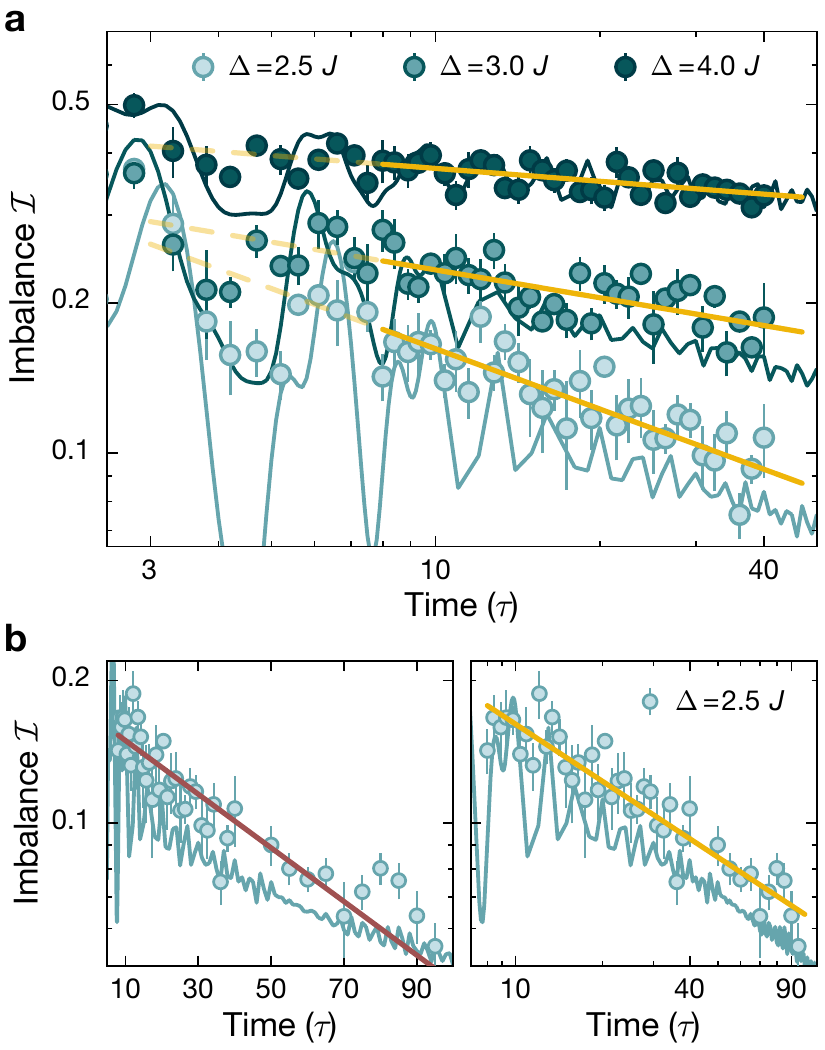}
	\caption{\textbf{Time evolution of the imbalance close to the MBL transition}: Decay of an initially prepared charge-density wave at a fixed interaction strength of $U = 4 \, J$. Points mark experimental data, averaged over six disorder phases $\phi$, with error bars indicating the uncertainty of the mean. The corresponding ED simulations for $S=20$ sites~\cite{SOMs} are indicated as solid lines. During the first three tunneling times (not shown), the imbalance quickly decays from its initial value close to one. During this initial decay, the imbalance shows strong oscillations, which cease after $\sim 8 \, \tau$. Thereafter, we observe a much slower further decay. \textbf{a)} Time traces for various disorder strengths with power-law fits. \textbf{b)} Long term decay at intermediate disorder strengths on a logarithmic y-axis with an exponential fit (left) and on a double-log plot with a power-law fit (right).}
	\label{traces_int_fig}
\end{figure}

\paragraph{\bf{Imbalance Time Traces.---}}
We monitor the dynamics in the interacting system via the time evolution of the imbalance for various disorder strengths above the noninteracting transition (see Fig.~\ref{traces_int_fig}a). The imbalance is shown on a log-log plot for times between $3-40 \, \tau$, which omits the rapid initial decay from its starting value close to one~\cite{Schreiber15}.  
After initial oscillations have ceased at around $8 \, \tau$, we observe slow relaxations of the imbalance, well reproduced by ED simulations (shown in Fig.~\ref{traces_int_fig}a, solid lines), which model our system on 20 sites~\cite{SOMs}. Upon increasing $\Delta$ this relaxation smoothly slows down until, for $\Delta \gtrsim 4 \, J$, the imbalance remains approximately constant, suggesting that the system becomes localized.

This dynamics in the quasiperiodic potential is reminiscent of the dynamics computed in numerical studies of true-random spin models~\cite{Luitz16}. In the true-random spin models, slow relaxation, which takes the form of power-laws, has been argued to result from rare, locally critical or insulating regions immersed in an otherwise thermal system~\cite{Vosk15,Potter15}.
However, the deterministic quasiperiodic potential in our system does not allow for such rare regions, raising the question of the microscopic mechanism and the functional form of the observed decays.

Fig.~\ref{traces_int_fig}b shows the time trace at $\Delta = 2.5 \, J$, to slightly longer evolution times of up to $100 \, \tau$. The data is presented on a lin-log (left panel) and a log-log (right panel) plot together with an exponential (red line) and a power-law (yellow line) fit to the experimental data. We find that the power-law fit describes the data slightly better than the exponential fit (see~\cite{SOMs} for fit residuals), a trend that is more pronounced in the numerical simulations. We attribute this difference to the background decay, present only in the experiment, that always contributes an exponential decay component, potentially altering the actual functional form. The numerical result is also consistent with a recent numerical study on spin models with quasiperiodic potentials~\cite{Lee17}, which also finds imbalance decays that are well described by power laws on intermediate time scales.
\\

\begin{figure}
	\centering
	\includegraphics[width=84mm]{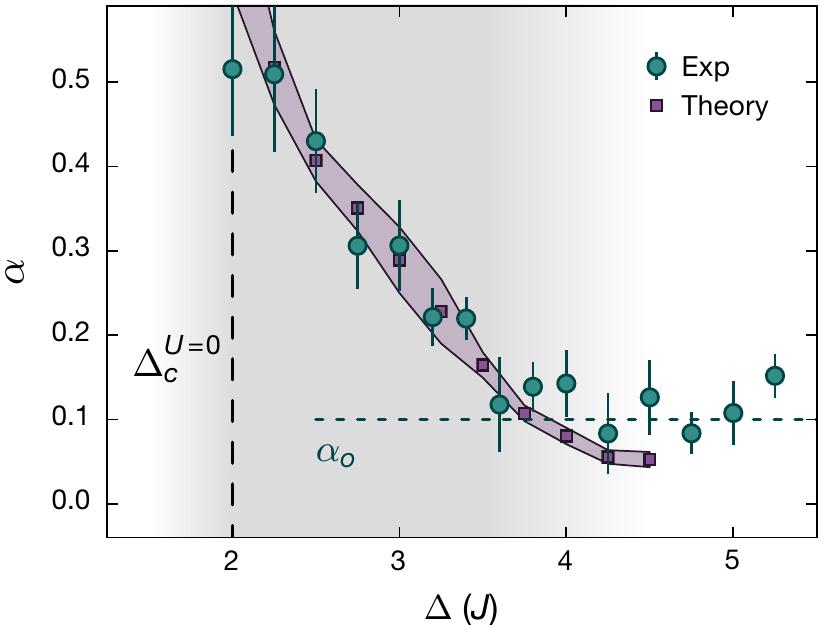}
	\caption{\textbf{Power-law exponent of imbalance decay:} Experimental and theoretical (ED, $S=20$, see~\cite{SOMs}) fitted exponents $\alpha$ as a function of disorder strength $\Delta$ at a fixed interaction strength of $U=4\,J$. Errorbars indicate the uncertainty of the fit to the experimental data. The purple shading denotes an estimate of the uncertainty on the simulated exponents based on finite size effects. For the largest disorder strengths, systematic errors due to finite time and size do not allow an accurate estimation of $\alpha$ and the actual uncertainty is likely underestimated~\cite{SOMs}. The gray shading marks the regime of slow dynamics as observed in Fig~\ref{imb_fig}. At large disorder strengths, the experimental value saturates at a nonzero offset $\alpha_o$, consistent with the independently observed background lifetime~\cite{SOMs,Bordia16,Lueschen16}. The finite value of $\alpha$ in ED for large disorder strength is likely caused by finite size effects. The corresponding exponents for the noninteracting data can be found in the supplemental material~\cite{SOMs}.}
	\label{exponents_fig}
\end{figure}

\paragraph{\bf{Relaxation Exponent.---}}
Motivated by the above analysis and the similarity to true-random systems~\cite{Luitz16}, we characterize the observed decays via power laws $\mathcal{I}(t) \sim t^{- \alpha}$. 
The exponents $\alpha$ are extracted using linear fits of $\log(\mathcal{I})$ versus $\log(t)$ between $8-40 \, \tau$ to the experimental data.
Fig.~\ref{exponents_fig} shows the experimental values in very good agreement with the results of ED simulations, where we choose a fitting range of $20-80 \, \tau$, as initial oscillations in the imbalance cease slower than in the experiment and affect the fitted exponent~\cite{SOMs}. 
Above the single-particle localization transition at $\Delta_{c}^{U=0} = 2 \, J$, we observe that $\alpha$ decreases monotonously until the experimental values saturate at a nonzero offset $\alpha_o$.
This offset is consistent with the expected effect of background decays in our system~\cite{SOMs,Bordia16,Lueschen16}, suggesting that $\alpha$ could indeed vanish in an isolated system. 
This suggests that the closed-system dynamics indeed smoothly changes from slow decays to a stationary finite imbalance at the MBL transition. 
We note though, that even in the MBL phase there may be a regime of slow, possibly logarithmic relaxation towards the stationary value of the imbalance~\cite{Mierzejewski16}, potentially contributing to a finite effective value of $\alpha_o$.

As in Ref.~\cite{Luitz16}, the exponents can be used to estimate the location of the MBL critical point as the disorder strength where the exponent becomes zero. In the experiment, however, this behavior is masked by the offset in the exponent resulting from the coupling to external baths.
As the effects of external baths on the power-law exponents (i.e.\ whether external decays result in a simple offset or a more complicated interplay) remain unclear, this prevents an accurate determination of $\Delta_c^\mathrm{MBL}$. However, the disorder strength where the exponents become compatible with the background decay does serve as a lower bound of $\Delta_c^\mathrm{MBL} \gtrsim 3.8 \pm 0.5 \, J$. The numerical results for small system sizes indicate that the actual critical disorder strength might be located at larger lattice depths and a simple linear extrapolation of the exponents gives a best guess for the critical disorder strength of $\Delta_c^\mathrm{MBL} \approx 4.3 \pm 0.5 \, J$. Previously performed DMRG for the localized phase suggest an upper bound for the MBL transition of $\Delta_c^\mathrm{MBL} \lesssim 5 \, J$~\cite{Bordia16_2}. For completeness, we also performed an equivalent analysis of the slow dynamics using exponential decays~\cite{SOMs}. While the individual fits are not quite as good as the power-law fits, similar bounds on the critical disorder strength can be obtained, further showing that the slowing down of the dynamics is a generic feature that captures the MBL transition in our system.

The lower bound for the transition exceeds the estimate of previous experimental work of $\Delta_c^\mathrm{MBL} \approx 2.5 \, J$~\cite{Schreiber15}. This value was extracted based on a finite-time measurement of the imbalance, a method that can become problematic in the presence of increasingly slow dynamics. The analysis based on the relaxation exponents given here takes into account the full dynamical evolution of the system and, hence, gives an improved estimate of the critical disorder strength.

The presented estimates of the critical point locate the MBL transition near the upper edge of the intermediate regime of slow dynamics in Fig.~\ref{imb_fig}. We note, that the upper edge of the noninteracting intermediate regime in Fig.~\ref{imb_fig} would slightly overestimate the known critical point of $\Delta_c^{U=0} = 2 \, J$~\cite{Aubry80}, as it neglects the initial dynamics on the localized side. Such a dynamics would, however, be much slower and possibly logarithmic in the MBL phase~\cite{Mierzejewski16}, and might, therefore, be masked by the background decay in the experiment.
\\

\paragraph{\bf{Discussion.---}}
We have experimentally observed a slow, interaction-induced relaxation dynamics close to the MBL phase transition in the interacting Aubry-Andr{\'e} model, in very good agreement with ED simulations. Specifically, we observe that the relaxation of an initial charge-density wave continuously slows down when approaching the MBL transition. On the experimentally accessible time scales, the decays are consistent with power laws whose exponents $\alpha$ smoothly vanish at the transition, thereby allowing for an estimation of the critical disorder strength based on the dynamics. 

As the dynamics observed in this experiment behave very similar to those found in numerical studies of true-random systems~\cite{Vosk15,Potter15,Agarwal15,Lev15,TorresHerrera15,Luitz16}, it is tempting to speculate whether the two systems share a common mechanism that underlies the slow dynamics. However, the Griffiths mechanism suggested to cause power-law dynamics in true-random systems~\cite{Vosk15,Potter15} cannot apply to quasiperiodic systems, as rare regions in the disorder pattern cannot exist in a deterministic potential. 
Given the wide regime of subdiffusive dynamics calculated in systems with true-random disorder~\cite{Luitz16,Znidaric16}, it is nonetheless possible that additional mechanisms are also at play in generating slow dynamics there.
It was suggested that one such mechanism could be strong local fluctuations in the initial state~\cite{Luitz16}, which are also present in our system. For instance, a region containing only one spin species would initially be noninteracting and, hence, insulating once the single-particle localization length is smaller than its size. The slow thermalization of such rare regions via their surroundings could give rise to power-law relaxation on intermediate time scales. On longer time scales, however, thermalization ultimately removes such regions and accelerates the imbalance relaxation. The melting of rare regions in the initial state might be further enhanced by the delocalized spin dynamics in our SU(2) symmetric system~\cite{Vasseur15,Prelovsek16,Protopopov16}.

Our results are consistent with two recent numerical studies on quasiperiodic systems that also find power-law decays of the imbalance on intermediate time scales~\cite{Lee17} and subdiffusive transport~\cite{BarLev17}. However, those properties have been found to exist also in the absence of randomness in the initial state, suggesting that rare regions in the initial state are at least not the sole cause of the slow dynamics. Instead, a further mechanism was proposed based on atypical transition rates between single-particle states~\cite{BarLev17}.

A similar mechanism was also suggested to explain the subdiffusive spreading of bosonic atoms in a quasiperiodic geometry observed in a previous experiment~\cite{Lucioni11}, which was performed in the absence of lattices along the orthogonal directions. Since this experiment was performed at a disorder strength where our system would appear localized, the dynamics likely emerged due to the bathlike effects of the delocalized orthogonal dimensions.

Our experimental and numerical results cannot distinguish which mechanism is relevant to the observed dynamics. The origin and exact functional shape of the slow dynamics pose an interesting open problem for future studies.
Experimentally, future studies could address the problem of a finite bath coupling via a systematic analysis of its effects~\cite{Lueschen16}, allowing for a further improvement in the determination of the transition point and potentially enabling access to the universal scaling regime.
\\

\begin{acknowledgments}
\paragraph{\bf{Acknowledgments.---}}
We acknowledge useful discussions with Eugene Demler, Michael Knap and Christophe Salomon. We acknowledge financial support by the European Commission (UQUAM, AQuS) and the Nanosystems Initiative Munich (NIM).
This work benefited from the support of the project THERMOLOC ANR-16-CE30-0023-02 of the French National Research Agency (ANR). ED simulations were performed using HPC resources from GENCI (Grants No. x2016050225 and x2017050225) and CALMIP (Grants No. 2016-P0677 and 2017-P0677). EA acknowledges the support from the Gyorgy chair of physics in UC Berkeley.
\end{acknowledgments}

\bibliography{CriticalSlowingDown}

\begin{thebibliography}{56}%
\makeatletter
\providecommand \@ifxundefined [1]{%
 \@ifx{#1\undefined}
}%
\providecommand \@ifnum [1]{%
 \ifnum #1\expandafter \@firstoftwo
 \else \expandafter \@secondoftwo
 \fi
}%
\providecommand \@ifx [1]{%
 \ifx #1\expandafter \@firstoftwo
 \else \expandafter \@secondoftwo
 \fi
}%
\providecommand \natexlab [1]{#1}%
\providecommand \enquote  [1]{``#1''}%
\providecommand \bibnamefont  [1]{#1}%
\providecommand \bibfnamefont [1]{#1}%
\providecommand \citenamefont [1]{#1}%
\providecommand \href@noop [0]{\@secondoftwo}%
\providecommand \href [0]{\begingroup \@sanitize@url \@href}%
\providecommand \@href[1]{\@@startlink{#1}\@@href}%
\providecommand \@@href[1]{\endgroup#1\@@endlink}%
\providecommand \@sanitize@url [0]{\catcode `\\12\catcode `\$12\catcode
  `\&12\catcode `\#12\catcode `\^12\catcode `\_12\catcode `\%12\relax}%
\providecommand \@@startlink[1]{}%
\providecommand \@@endlink[0]{}%
\providecommand \url  [0]{\begingroup\@sanitize@url \@url }%
\providecommand \@url [1]{\endgroup\@href {#1}{\urlprefix }}%
\providecommand \urlprefix  [0]{URL }%
\providecommand \Eprint [0]{\href }%
\providecommand \doibase [0]{http://dx.doi.org/}%
\providecommand \selectlanguage [0]{\@gobble}%
\providecommand \bibinfo  [0]{\@secondoftwo}%
\providecommand \bibfield  [0]{\@secondoftwo}%
\providecommand \translation [1]{[#1]}%
\providecommand \BibitemOpen [0]{}%
\providecommand \bibitemStop [0]{}%
\providecommand \bibitemNoStop [0]{.\EOS\space}%
\providecommand \EOS [0]{\spacefactor3000\relax}%
\providecommand \BibitemShut  [1]{\csname bibitem#1\endcsname}%
\let\auto@bib@innerbib\@empty
%</preamble>
\bibitem [{\citenamefont {Basko}\ \emph {et~al.}(2006)\citenamefont {Basko},
  \citenamefont {Aleiner},\ and\ \citenamefont {Altshuler}}]{Basko06}%
  \BibitemOpen
  \bibfield  {author} {\bibinfo {author} {\bibfnamefont {D.M.}\ \bibnamefont
  {Basko}}, \bibinfo {author} {\bibfnamefont {I.~L.}\ \bibnamefont {Aleiner}},
  \ and\ \bibinfo {author} {\bibfnamefont {B.~L.}\ \bibnamefont {Altshuler}},\
  }\bibfield  {title} {\enquote {\bibinfo {title} {Metal-insulator transition
  in a weakly interacting many-electron system with localized single-particle
  states},}\ }\href {http://dx.doi.org/10.1016/j.aop.2005.11.014} {\bibfield
  {journal} {\bibinfo  {journal} {Ann. Phys.}\ }\textbf {\bibinfo {volume}
  {321}},\ \bibinfo {pages} {1126} (\bibinfo {year} {2006})}\BibitemShut
  {NoStop}%
\bibitem [{\citenamefont {Gornyi}\ \emph {et~al.}(2005)\citenamefont {Gornyi},
  \citenamefont {Mirlin},\ and\ \citenamefont {Polyakov}}]{Polyakov05}%
  \BibitemOpen
  \bibfield  {author} {\bibinfo {author} {\bibfnamefont {I.~V.}\ \bibnamefont
  {Gornyi}}, \bibinfo {author} {\bibfnamefont {A.~D.}\ \bibnamefont {Mirlin}},
  \ and\ \bibinfo {author} {\bibfnamefont {D.~G.}\ \bibnamefont {Polyakov}},\
  }\bibfield  {title} {\enquote {\bibinfo {title} {Interacting electrons in
  disordered wires: Anderson localization and low-$t$ transport},}\ }\href
  {\doibase 10.1103/PhysRevLett.95.206603} {\bibfield  {journal} {\bibinfo
  {journal} {Phys. Rev. Lett.}\ }\textbf {\bibinfo {volume} {95}},\ \bibinfo
  {pages} {206603} (\bibinfo {year} {2005})}\BibitemShut {NoStop}%
\bibitem [{\citenamefont {Imbrie}(2016)}]{Imbrie16}%
  \BibitemOpen
  \bibfield  {author} {\bibinfo {author} {\bibfnamefont {John~Z.}\ \bibnamefont
  {Imbrie}},\ }\bibfield  {title} {\enquote {\bibinfo {title} {On many-body
  localization for quantum spin chains},}\ }\href {\doibase
  10.1007/s10955-016-1508-x} {\bibfield  {journal} {\bibinfo  {journal} {J.
  Stat. Phys.}\ }\textbf {\bibinfo {volume} {163}},\ \bibinfo {pages}
  {998--1048} (\bibinfo {year} {2016})}\BibitemShut {NoStop}%
\bibitem [{\citenamefont {Oganesyan}\ and\ \citenamefont
  {Huse}(2007)}]{Oganesyan05}%
  \BibitemOpen
  \bibfield  {author} {\bibinfo {author} {\bibfnamefont {V.}~\bibnamefont
  {Oganesyan}}\ and\ \bibinfo {author} {\bibfnamefont {D.~A.}\ \bibnamefont
  {Huse}},\ }\bibfield  {title} {\enquote {\bibinfo {title} {Localization of
  interacting fermions at high temperature},}\ }\href {\doibase
  10.1103/PhysRevB.75.155111} {\bibfield  {journal} {\bibinfo  {journal} {Phys.
  Rev. B}\ }\textbf {\bibinfo {volume} {75}},\ \bibinfo {pages} {155111}
  (\bibinfo {year} {2007})}\BibitemShut {NoStop}%
\bibitem [{\citenamefont {Pal}\ and\ \citenamefont {Huse}(2010)}]{Pal10}%
  \BibitemOpen
  \bibfield  {author} {\bibinfo {author} {\bibfnamefont {A.}~\bibnamefont
  {Pal}}\ and\ \bibinfo {author} {\bibfnamefont {D.~A.}\ \bibnamefont {Huse}},\
  }\bibfield  {title} {\enquote {\bibinfo {title} {Many-body localization phase
  transition},}\ }\href {\doibase 10.1103/PhysRevB.82.174411} {\bibfield
  {journal} {\bibinfo  {journal} {Phys. Rev. B}\ }\textbf {\bibinfo {volume}
  {82}},\ \bibinfo {pages} {174411} (\bibinfo {year} {2010})}\BibitemShut
  {NoStop}%
\bibitem [{\citenamefont {Vosk}\ and\ \citenamefont {Altman}(2013)}]{Vosk13}%
  \BibitemOpen
  \bibfield  {author} {\bibinfo {author} {\bibfnamefont {R.}~\bibnamefont
  {Vosk}}\ and\ \bibinfo {author} {\bibfnamefont {E.}~\bibnamefont {Altman}},\
  }\bibfield  {title} {\enquote {\bibinfo {title} {Many-body localization in
  one dimension as a dynamical renormalization group fixed point},}\ }\href
  {\doibase 10.1103/PhysRevLett.110.067204} {\bibfield  {journal} {\bibinfo
  {journal} {Phys. Rev. Lett.}\ }\textbf {\bibinfo {volume} {110}},\ \bibinfo
  {pages} {067204} (\bibinfo {year} {2013})}\BibitemShut {NoStop}%
\bibitem [{\citenamefont {Serbyn}\ \emph {et~al.}(2014)\citenamefont {Serbyn},
  \citenamefont {Knap}, \citenamefont {Gopalakrishnan}, \citenamefont
  {Papi\ifmmode~\acute{c}\else \'{c}\fi{}}, \citenamefont {Yao}, \citenamefont
  {Laumann}, \citenamefont {Abanin}, \citenamefont {Lukin},\ and\ \citenamefont
  {Demler}}]{Serbyn14}%
  \BibitemOpen
  \bibfield  {author} {\bibinfo {author} {\bibfnamefont {M.}~\bibnamefont
  {Serbyn}}, \bibinfo {author} {\bibfnamefont {M.}~\bibnamefont {Knap}},
  \bibinfo {author} {\bibfnamefont {S.}~\bibnamefont {Gopalakrishnan}},
  \bibinfo {author} {\bibfnamefont {Z.}~\bibnamefont
  {Papi\ifmmode~\acute{c}\else \'{c}\fi{}}}, \bibinfo {author} {\bibfnamefont
  {N.~Y.}\ \bibnamefont {Yao}}, \bibinfo {author} {\bibfnamefont {C.~R.}\
  \bibnamefont {Laumann}}, \bibinfo {author} {\bibfnamefont {D.~A.}\
  \bibnamefont {Abanin}}, \bibinfo {author} {\bibfnamefont {M.~D.}\
  \bibnamefont {Lukin}}, \ and\ \bibinfo {author} {\bibfnamefont {E.~A.}\
  \bibnamefont {Demler}},\ }\bibfield  {title} {\enquote {\bibinfo {title}
  {Interferometric probes of many-body localization},}\ }\href {\doibase
  10.1103/PhysRevLett.113.147204} {\bibfield  {journal} {\bibinfo  {journal}
  {Phys. Rev. Lett.}\ }\textbf {\bibinfo {volume} {113}},\ \bibinfo {pages}
  {147204} (\bibinfo {year} {2014})}\BibitemShut {NoStop}%
\bibitem [{\citenamefont {Michal}\ \emph {et~al.}(2016)\citenamefont {Michal},
  \citenamefont {Aleiner}, \citenamefont {Altshuler},\ and\ \citenamefont
  {Shlyapnikov}}]{Shlyapnikov15}%
  \BibitemOpen
  \bibfield  {author} {\bibinfo {author} {\bibfnamefont {V.~P.}\ \bibnamefont
  {Michal}}, \bibinfo {author} {\bibfnamefont {I.~L.}\ \bibnamefont {Aleiner}},
  \bibinfo {author} {\bibfnamefont {B.~L.}\ \bibnamefont {Altshuler}}, \ and\
  \bibinfo {author} {\bibfnamefont {G.~V.}\ \bibnamefont {Shlyapnikov}},\
  }\bibfield  {title} {\enquote {\bibinfo {title} {Finite-temperature
  fluid–insulator transition of strongly interacting 1d disordered bosons},}\
  }\href {\doibase 10.1073/pnas.1606908113} {\bibfield  {journal} {\bibinfo
  {journal} {Proc. Natl. Acad. Sci. U.S.A.}\ }\textbf {\bibinfo {volume}
  {113}},\ \bibinfo {pages} {E4455--E4459} (\bibinfo {year}
  {2016})}\BibitemShut {NoStop}%
\bibitem [{\citenamefont {Nandkishore}\ and\ \citenamefont
  {Huse}(2015)}]{Nandkishore15}%
  \BibitemOpen
  \bibfield  {author} {\bibinfo {author} {\bibfnamefont {R.}~\bibnamefont
  {Nandkishore}}\ and\ \bibinfo {author} {\bibfnamefont {D.~A.}\ \bibnamefont
  {Huse}},\ }\bibfield  {title} {\enquote {\bibinfo {title} {Many-body
  localization and thermalization in quantum statistical mechanics},}\ }\href
  {\doibase 10.1146/annurev-conmatphys-031214-014726} {\bibfield  {journal}
  {\bibinfo  {journal} {Annu. Rev. Condens. Matter Phys.}\ }\textbf {\bibinfo
  {volume} {6}},\ \bibinfo {pages} {15--38} (\bibinfo {year}
  {2015})}\BibitemShut {NoStop}%
\bibitem [{\citenamefont {Altman}\ and\ \citenamefont {Vosk}(2015)}]{Altman15}%
  \BibitemOpen
  \bibfield  {author} {\bibinfo {author} {\bibfnamefont {E.}~\bibnamefont
  {Altman}}\ and\ \bibinfo {author} {\bibfnamefont {R.}~\bibnamefont {Vosk}},\
  }\bibfield  {title} {\enquote {\bibinfo {title} {Universal dynamics and
  renormalization in many-body-localized systems},}\ }\href
  {http://dx.doi.org/10.1146/annurev-conmatphys-031214-014701} {\bibfield
  {journal} {\bibinfo  {journal} {Annu. Rev. Condens. Matter Phys.}\ }\textbf
  {\bibinfo {volume} {6}},\ \bibinfo {pages} {383--409} (\bibinfo {year}
  {2015})}\BibitemShut {NoStop}%
\bibitem [{\citenamefont {Bahri}\ \emph {et~al.}(2015)\citenamefont {Bahri},
  \citenamefont {Vosk}, \citenamefont {Altman},\ and\ \citenamefont
  {Vishwanath}}]{Bahri15}%
  \BibitemOpen
  \bibfield  {author} {\bibinfo {author} {\bibfnamefont {Y.}~\bibnamefont
  {Bahri}}, \bibinfo {author} {\bibfnamefont {R.}~\bibnamefont {Vosk}},
  \bibinfo {author} {\bibfnamefont {E.}~\bibnamefont {Altman}}, \ and\ \bibinfo
  {author} {\bibfnamefont {A.}~\bibnamefont {Vishwanath}},\ }\bibfield  {title}
  {\enquote {\bibinfo {title} {Localization and topology protected quantum
  coherence at the edge of hot matter},}\ }\href
  {http://dx.doi.org/10.1038/ncomms8341} {\bibfield  {journal} {\bibinfo
  {journal} {Nat. Commun.}\ }\textbf {\bibinfo {volume} {6}},\ \bibinfo {pages}
  {7341} (\bibinfo {year} {2015})}\BibitemShut {NoStop}%
\bibitem [{\citenamefont {Ovadia}\ \emph {et~al.}(2015)\citenamefont {Ovadia},
  \citenamefont {Kalok}, \citenamefont {Tamir}, \citenamefont {Mitra},
  \citenamefont {Sac{\'e}p{\'e}},\ and\ \citenamefont {Shahar}}]{Ovadia15}%
  \BibitemOpen
  \bibfield  {author} {\bibinfo {author} {\bibfnamefont {M.}~\bibnamefont
  {Ovadia}}, \bibinfo {author} {\bibfnamefont {D.}~\bibnamefont {Kalok}},
  \bibinfo {author} {\bibfnamefont {I.}~\bibnamefont {Tamir}}, \bibinfo
  {author} {\bibfnamefont {S.}~\bibnamefont {Mitra}}, \bibinfo {author}
  {\bibfnamefont {B.}~\bibnamefont {Sac{\'e}p{\'e}}}, \ and\ \bibinfo {author}
  {\bibfnamefont {D.}~\bibnamefont {Shahar}},\ }\bibfield  {title} {\enquote
  {\bibinfo {title} {Evidence for a finite-temperature insulator},}\ }\href
  {http://dx.doi.org/10.1038/srep13503} {\bibfield  {journal} {\bibinfo
  {journal} {Sci. Rep.}\ }\textbf {\bibinfo {volume} {5}},\ \bibinfo {pages}
  {13503} (\bibinfo {year} {2015})}\BibitemShut {NoStop}%
\bibitem [{\citenamefont {Schreiber}\ \emph {et~al.}(2015)\citenamefont
  {Schreiber}, \citenamefont {Hodgman}, \citenamefont {Bordia}, \citenamefont
  {L{\"u}schen}, \citenamefont {Fischer}, \citenamefont {Vosk}, \citenamefont
  {Altman}, \citenamefont {Schneider},\ and\ \citenamefont
  {Bloch}}]{Schreiber15}%
  \BibitemOpen
  \bibfield  {author} {\bibinfo {author} {\bibfnamefont {M.}~\bibnamefont
  {Schreiber}}, \bibinfo {author} {\bibfnamefont {S.~S.}\ \bibnamefont
  {Hodgman}}, \bibinfo {author} {\bibfnamefont {P.}~\bibnamefont {Bordia}},
  \bibinfo {author} {\bibfnamefont {H.~P.}\ \bibnamefont {L{\"u}schen}},
  \bibinfo {author} {\bibfnamefont {M.~H.}\ \bibnamefont {Fischer}}, \bibinfo
  {author} {\bibfnamefont {R.}~\bibnamefont {Vosk}}, \bibinfo {author}
  {\bibfnamefont {E.}~\bibnamefont {Altman}}, \bibinfo {author} {\bibfnamefont
  {U.}~\bibnamefont {Schneider}}, \ and\ \bibinfo {author} {\bibfnamefont
  {I.}~\bibnamefont {Bloch}},\ }\bibfield  {title} {\enquote {\bibinfo {title}
  {Observation of many-body localization of interacting fermions in a
  quasirandom optical lattice},}\ }\href {\doibase 10.1126/science.aaa7432}
  {\bibfield  {journal} {\bibinfo  {journal} {Science}\ }\textbf {\bibinfo
  {volume} {349}},\ \bibinfo {pages} {842--845} (\bibinfo {year}
  {2015})}\BibitemShut {NoStop}%
\bibitem [{\citenamefont {Kondov}\ \emph {et~al.}(2015)\citenamefont {Kondov},
  \citenamefont {McGehee}, \citenamefont {Xu},\ and\ \citenamefont
  {DeMarco}}]{Kondov15}%
  \BibitemOpen
  \bibfield  {author} {\bibinfo {author} {\bibfnamefont {S.~S.}\ \bibnamefont
  {Kondov}}, \bibinfo {author} {\bibfnamefont {W.~R.}\ \bibnamefont {McGehee}},
  \bibinfo {author} {\bibfnamefont {W.}~\bibnamefont {Xu}}, \ and\ \bibinfo
  {author} {\bibfnamefont {B.}~\bibnamefont {DeMarco}},\ }\bibfield  {title}
  {\enquote {\bibinfo {title} {Disorder-induced localization in a strongly
  correlated atomic hubbard gas},}\ }\href {\doibase
  10.1103/PhysRevLett.114.083002} {\bibfield  {journal} {\bibinfo  {journal}
  {Phys. Rev. Lett.}\ }\textbf {\bibinfo {volume} {114}},\ \bibinfo {pages}
  {083002} (\bibinfo {year} {2015})}\BibitemShut {NoStop}%
\bibitem [{\citenamefont {Smith}\ \emph {et~al.}(2016)\citenamefont {Smith},
  \citenamefont {Lee}, \citenamefont {Richerme}, \citenamefont {Neyenhuis},
  \citenamefont {Hess}, \citenamefont {Hauke}, \citenamefont {Heyl},
  \citenamefont {Huse},\ and\ \citenamefont {Monroe}}]{Smith15}%
  \BibitemOpen
  \bibfield  {author} {\bibinfo {author} {\bibfnamefont {J.}~\bibnamefont
  {Smith}}, \bibinfo {author} {\bibfnamefont {A.}~\bibnamefont {Lee}}, \bibinfo
  {author} {\bibfnamefont {P.}~\bibnamefont {Richerme}}, \bibinfo {author}
  {\bibfnamefont {B.}~\bibnamefont {Neyenhuis}}, \bibinfo {author}
  {\bibfnamefont {P.~W.}\ \bibnamefont {Hess}}, \bibinfo {author}
  {\bibfnamefont {P.}~\bibnamefont {Hauke}}, \bibinfo {author} {\bibfnamefont
  {M.}~\bibnamefont {Heyl}}, \bibinfo {author} {\bibfnamefont {D.~A.}\
  \bibnamefont {Huse}}, \ and\ \bibinfo {author} {\bibfnamefont
  {C.}~\bibnamefont {Monroe}},\ }\bibfield  {title} {\enquote {\bibinfo {title}
  {Many-body localization in a quantum simulator with programmable random
  disorder},}\ }\href {http://dx.doi.org/10.1038/nphys3783} {\bibfield
  {journal} {\bibinfo  {journal} {Nat. Phys.}\ }\textbf {\bibinfo {volume}
  {12}},\ \bibinfo {pages} {907--911} (\bibinfo {year} {2016})}\BibitemShut
  {NoStop}%
\bibitem [{\citenamefont {Choi}\ \emph {et~al.}(2016)\citenamefont {Choi},
  \citenamefont {Hild}, \citenamefont {Zeiher}, \citenamefont {Schau{\ss}},
  \citenamefont {Rubio-Abadal}, \citenamefont {Yefsah}, \citenamefont
  {Khemani}, \citenamefont {Huse}, \citenamefont {Bloch},\ and\ \citenamefont
  {Gross}}]{Choi16}%
  \BibitemOpen
  \bibfield  {author} {\bibinfo {author} {\bibfnamefont {J.-y.}\ \bibnamefont
  {Choi}}, \bibinfo {author} {\bibfnamefont {S.}~\bibnamefont {Hild}}, \bibinfo
  {author} {\bibfnamefont {J.}~\bibnamefont {Zeiher}}, \bibinfo {author}
  {\bibfnamefont {P.}~\bibnamefont {Schau{\ss}}}, \bibinfo {author}
  {\bibfnamefont {A.}~\bibnamefont {Rubio-Abadal}}, \bibinfo {author}
  {\bibfnamefont {T.}~\bibnamefont {Yefsah}}, \bibinfo {author} {\bibfnamefont
  {V.}~\bibnamefont {Khemani}}, \bibinfo {author} {\bibfnamefont {D.~A.}\
  \bibnamefont {Huse}}, \bibinfo {author} {\bibfnamefont {I.}~\bibnamefont
  {Bloch}}, \ and\ \bibinfo {author} {\bibfnamefont {C.}~\bibnamefont
  {Gross}},\ }\bibfield  {title} {\enquote {\bibinfo {title} {Exploring the
  many-body localization transition in two dimensions},}\ }\href {\doibase
  10.1126/science.aaf8834} {\bibfield  {journal} {\bibinfo  {journal}
  {Science}\ }\textbf {\bibinfo {volume} {352}},\ \bibinfo {pages} {1547--1552}
  (\bibinfo {year} {2016})}\BibitemShut {NoStop}%
\bibitem [{\citenamefont {Roati}\ \emph {et~al.}(2008)\citenamefont {Roati},
  \citenamefont {D'Errico}, \citenamefont {Fallani}, \citenamefont {Fattori},
  \citenamefont {Fort}, \citenamefont {Zaccanti}, \citenamefont {Modugno},
  \citenamefont {Modugno},\ and\ \citenamefont {Inguscio}}]{Roati08}%
  \BibitemOpen
  \bibfield  {author} {\bibinfo {author} {\bibfnamefont {G.}~\bibnamefont
  {Roati}}, \bibinfo {author} {\bibfnamefont {C.}~\bibnamefont {D'Errico}},
  \bibinfo {author} {\bibfnamefont {L.}~\bibnamefont {Fallani}}, \bibinfo
  {author} {\bibfnamefont {M.}~\bibnamefont {Fattori}}, \bibinfo {author}
  {\bibfnamefont {C.}~\bibnamefont {Fort}}, \bibinfo {author} {\bibfnamefont
  {M.}~\bibnamefont {Zaccanti}}, \bibinfo {author} {\bibfnamefont
  {G.}~\bibnamefont {Modugno}}, \bibinfo {author} {\bibfnamefont
  {M.}~\bibnamefont {Modugno}}, \ and\ \bibinfo {author} {\bibfnamefont
  {M.}~\bibnamefont {Inguscio}},\ }\bibfield  {title} {\enquote {\bibinfo
  {title} {Anderson localization of a non-interacting bose-einstein
  condensate},}\ }\href {\doibase 10.1038/nature07071} {\bibfield  {journal}
  {\bibinfo  {journal} {Nature}\ }\textbf {\bibinfo {volume} {453}},\ \bibinfo
  {pages} {895--898} (\bibinfo {year} {2008})}\BibitemShut {NoStop}%
\bibitem [{\citenamefont {Iyer}\ \emph {et~al.}(2013)\citenamefont {Iyer},
  \citenamefont {Oganesyan}, \citenamefont {Refael},\ and\ \citenamefont
  {Huse}}]{Iyer13}%
  \BibitemOpen
  \bibfield  {author} {\bibinfo {author} {\bibfnamefont {S.}~\bibnamefont
  {Iyer}}, \bibinfo {author} {\bibfnamefont {V.}~\bibnamefont {Oganesyan}},
  \bibinfo {author} {\bibfnamefont {G.}~\bibnamefont {Refael}}, \ and\ \bibinfo
  {author} {\bibfnamefont {D.~A.}\ \bibnamefont {Huse}},\ }\bibfield  {title}
  {\enquote {\bibinfo {title} {Many-body localization in a quasiperiodic
  system},}\ }\href {\doibase 10.1103/PhysRevB.87.134202} {\bibfield  {journal}
  {\bibinfo  {journal} {Phys. Rev. B}\ }\textbf {\bibinfo {volume} {87}},\
  \bibinfo {pages} {134202} (\bibinfo {year} {2013})}\BibitemShut {NoStop}%
\bibitem [{\citenamefont {Deutsch}(1991)}]{Deutsch91}%
  \BibitemOpen
  \bibfield  {author} {\bibinfo {author} {\bibfnamefont {J.~M.}\ \bibnamefont
  {Deutsch}},\ }\bibfield  {title} {\enquote {\bibinfo {title} {Quantum
  statistical mechanics in a closed system},}\ }\href {\doibase
  10.1103/PhysRevA.43.2046} {\bibfield  {journal} {\bibinfo  {journal} {Phys.
  Rev. A}\ }\textbf {\bibinfo {volume} {43}},\ \bibinfo {pages} {2046--2049}
  (\bibinfo {year} {1991})}\BibitemShut {NoStop}%
\bibitem [{\citenamefont {Srednicki}(1994)}]{Sred94}%
  \BibitemOpen
  \bibfield  {author} {\bibinfo {author} {\bibfnamefont {M.}~\bibnamefont
  {Srednicki}},\ }\bibfield  {title} {\enquote {\bibinfo {title} {Chaos and
  quantum thermalization},}\ }\href {\doibase 10.1103/PhysRevE.50.888}
  {\bibfield  {journal} {\bibinfo  {journal} {Phys. Rev. E}\ }\textbf {\bibinfo
  {volume} {50}},\ \bibinfo {pages} {888--901} (\bibinfo {year}
  {1994})}\BibitemShut {NoStop}%
\bibitem [{\citenamefont {Rigol}\ \emph {et~al.}(2008)\citenamefont {Rigol},
  \citenamefont {Dunjko},\ and\ \citenamefont {Olshanii}}]{Rigol08}%
  \BibitemOpen
  \bibfield  {author} {\bibinfo {author} {\bibfnamefont {M.}~\bibnamefont
  {Rigol}}, \bibinfo {author} {\bibfnamefont {V.}~\bibnamefont {Dunjko}}, \
  and\ \bibinfo {author} {\bibfnamefont {M.}~\bibnamefont {Olshanii}},\
  }\bibfield  {title} {\enquote {\bibinfo {title} {Thermalization and its
  mechanism for generic isolated quantum systems},}\ }\href
  {http://dx.doi.org/10.1038/nature06838} {\bibfield  {journal} {\bibinfo
  {journal} {Nature}\ }\textbf {\bibinfo {volume} {452}} (\bibinfo {year}
  {2008})}\BibitemShut {NoStop}%
\bibitem [{\citenamefont {Parameswaran}\ \emph {et~al.}(2017)\citenamefont
  {Parameswaran}, \citenamefont {Potter},\ and\ \citenamefont
  {Vasseur}}]{Vasseur16}%
  \BibitemOpen
  \bibfield  {author} {\bibinfo {author} {\bibfnamefont {S.~A.}\ \bibnamefont
  {Parameswaran}}, \bibinfo {author} {\bibfnamefont {A.~C.}\ \bibnamefont
  {Potter}}, \ and\ \bibinfo {author} {\bibfnamefont {R.}~\bibnamefont
  {Vasseur}},\ }\bibfield  {title} {\enquote {\bibinfo {title} {Eigenstate
  phase transitions and the emergence of universal dynamics in highly excited
  states},}\ }\href {\doibase 10.1002/andp.201600302} {\bibfield  {journal}
  {\bibinfo  {journal} {Ann. Phys.}\ }\textbf {\bibinfo {volume} {529}},\
  \bibinfo {pages} {1600302} (\bibinfo {year} {2017})}\BibitemShut {NoStop}%
\bibitem [{\citenamefont {Luitz}\ and\ \citenamefont {Lev}(2017)}]{Luitz16_3}%
  \BibitemOpen
  \bibfield  {author} {\bibinfo {author} {\bibfnamefont {D.~J.}\ \bibnamefont
  {Luitz}}\ and\ \bibinfo {author} {\bibfnamefont {Y.~Bar}\ \bibnamefont
  {Lev}},\ }\bibfield  {title} {\enquote {\bibinfo {title} {The ergodic side of
  the many-body localization transition},}\ }\href {\doibase
  10.1002/andp.201600350} {\bibfield  {journal} {\bibinfo  {journal} {Ann.
  Phys.}\ }\textbf {\bibinfo {volume} {529}},\ \bibinfo {pages} {1600350}
  (\bibinfo {year} {2017})}\BibitemShut {NoStop}%
\bibitem [{\citenamefont {Agarwal}\ \emph {et~al.}(2017)\citenamefont
  {Agarwal}, \citenamefont {Altman}, \citenamefont {Demler}, \citenamefont
  {Gopalakrishnan}, \citenamefont {Huse},\ and\ \citenamefont
  {Knap}}]{Agarwal16}%
  \BibitemOpen
  \bibfield  {author} {\bibinfo {author} {\bibfnamefont {K.}~\bibnamefont
  {Agarwal}}, \bibinfo {author} {\bibfnamefont {E.}~\bibnamefont {Altman}},
  \bibinfo {author} {\bibfnamefont {E.}~\bibnamefont {Demler}}, \bibinfo
  {author} {\bibfnamefont {S.}~\bibnamefont {Gopalakrishnan}}, \bibinfo
  {author} {\bibfnamefont {D.~A.}\ \bibnamefont {Huse}}, \ and\ \bibinfo
  {author} {\bibfnamefont {M.}~\bibnamefont {Knap}},\ }\bibfield  {title}
  {\enquote {\bibinfo {title} {Rare-region effects and dynamics near the
  many-body localization transition},}\ }\href {\doibase
  10.1002/andp.201600326} {\bibfield  {journal} {\bibinfo  {journal} {Annalen
  der Physik}\ }\textbf {\bibinfo {volume} {529}},\ \bibinfo {pages} {1600326}
  (\bibinfo {year} {2017})}\BibitemShut {NoStop}%
\bibitem [{\citenamefont {Sachdev}(2001)}]{Sachdev_book}%
  \BibitemOpen
  \bibfield  {author} {\bibinfo {author} {\bibfnamefont {S.}~\bibnamefont
  {Sachdev}},\ }\href@noop {} {\emph {\bibinfo {title} {Quantum Phase
  Transitions}}}\ (\bibinfo  {publisher} {Cambridge University Press},\
  \bibinfo {year} {2001})\BibitemShut {NoStop}%
\bibitem [{\citenamefont {Luitz}\ and\ \citenamefont
  {Bar~Lev}(2016)}]{Luitz16_2}%
  \BibitemOpen
  \bibfield  {author} {\bibinfo {author} {\bibfnamefont {D.~J.}\ \bibnamefont
  {Luitz}}\ and\ \bibinfo {author} {\bibfnamefont {Y.}~\bibnamefont
  {Bar~Lev}},\ }\bibfield  {title} {\enquote {\bibinfo {title} {Anomalous
  thermalization in ergodic systems},}\ }\href {\doibase
  10.1103/PhysRevLett.117.170404} {\bibfield  {journal} {\bibinfo  {journal}
  {Phys. Rev. Lett.}\ }\textbf {\bibinfo {volume} {117}},\ \bibinfo {pages}
  {170404} (\bibinfo {year} {2016})}\BibitemShut {NoStop}%
\bibitem [{\citenamefont {Khemani}\ \emph
  {et~al.}(2017{\natexlab{a}})\citenamefont {Khemani}, \citenamefont {Lim},
  \citenamefont {Sheng},\ and\ \citenamefont {Huse}}]{Khemani16}%
  \BibitemOpen
  \bibfield  {author} {\bibinfo {author} {\bibfnamefont {V.}~\bibnamefont
  {Khemani}}, \bibinfo {author} {\bibfnamefont {S.~P.}\ \bibnamefont {Lim}},
  \bibinfo {author} {\bibfnamefont {D.~N.}\ \bibnamefont {Sheng}}, \ and\
  \bibinfo {author} {\bibfnamefont {D.~A.}\ \bibnamefont {Huse}},\ }\bibfield
  {title} {\enquote {\bibinfo {title} {Critical properties of the many-body
  localization transition},}\ }\href {\doibase 10.1103/PhysRevX.7.021013}
  {\bibfield  {journal} {\bibinfo  {journal} {Phys. Rev. X}\ }\textbf {\bibinfo
  {volume} {7}},\ \bibinfo {pages} {021013} (\bibinfo {year}
  {2017}{\natexlab{a}})}\BibitemShut {NoStop}%
\bibitem [{\citenamefont {Vosk}\ \emph {et~al.}(2015)\citenamefont {Vosk},
  \citenamefont {Huse},\ and\ \citenamefont {Altman}}]{Vosk15}%
  \BibitemOpen
  \bibfield  {author} {\bibinfo {author} {\bibfnamefont {R.}~\bibnamefont
  {Vosk}}, \bibinfo {author} {\bibfnamefont {D.~A.}\ \bibnamefont {Huse}}, \
  and\ \bibinfo {author} {\bibfnamefont {E.}~\bibnamefont {Altman}},\
  }\bibfield  {title} {\enquote {\bibinfo {title} {Theory of the many-body
  localization transition in one-dimensional systems},}\ }\href {\doibase
  10.1103/PhysRevX.5.031032} {\bibfield  {journal} {\bibinfo  {journal} {Phys.
  Rev. X}\ }\textbf {\bibinfo {volume} {5}},\ \bibinfo {pages} {031032}
  (\bibinfo {year} {2015})}\BibitemShut {NoStop}%
\bibitem [{\citenamefont {Potter}\ \emph {et~al.}(2015)\citenamefont {Potter},
  \citenamefont {Vasseur},\ and\ \citenamefont {Parameswaran}}]{Potter15}%
  \BibitemOpen
  \bibfield  {author} {\bibinfo {author} {\bibfnamefont {A.~C.}\ \bibnamefont
  {Potter}}, \bibinfo {author} {\bibfnamefont {R.}~\bibnamefont {Vasseur}}, \
  and\ \bibinfo {author} {\bibfnamefont {S.~A.}\ \bibnamefont {Parameswaran}},\
  }\bibfield  {title} {\enquote {\bibinfo {title} {Universal properties of
  many-body delocalization transitions},}\ }\href {\doibase
  10.1103/PhysRevX.5.031033} {\bibfield  {journal} {\bibinfo  {journal} {Phys.
  Rev. X}\ }\textbf {\bibinfo {volume} {5}},\ \bibinfo {pages} {031033}
  (\bibinfo {year} {2015})}\BibitemShut {NoStop}%
\bibitem [{\citenamefont {Griffiths}(1969)}]{Griffiths69}%
  \BibitemOpen
  \bibfield  {author} {\bibinfo {author} {\bibfnamefont {R.~B.}\ \bibnamefont
  {Griffiths}},\ }\bibfield  {title} {\enquote {\bibinfo {title} {Nonanalytic
  behavior above the critical point in a random ising ferromagnet},}\ }\href
  {\doibase 10.1103/PhysRevLett.23.17} {\bibfield  {journal} {\bibinfo
  {journal} {Phys. Rev. Lett.}\ }\textbf {\bibinfo {volume} {23}},\ \bibinfo
  {pages} {17--19} (\bibinfo {year} {1969})}\BibitemShut {NoStop}%
\bibitem [{\citenamefont {Agarwal}\ \emph {et~al.}(2015)\citenamefont
  {Agarwal}, \citenamefont {Gopalakrishnan}, \citenamefont {Knap},
  \citenamefont {M\"uller},\ and\ \citenamefont {Demler}}]{Agarwal15}%
  \BibitemOpen
  \bibfield  {author} {\bibinfo {author} {\bibfnamefont {K.}~\bibnamefont
  {Agarwal}}, \bibinfo {author} {\bibfnamefont {S.}~\bibnamefont
  {Gopalakrishnan}}, \bibinfo {author} {\bibfnamefont {M.}~\bibnamefont
  {Knap}}, \bibinfo {author} {\bibfnamefont {M.}~\bibnamefont {M\"uller}}, \
  and\ \bibinfo {author} {\bibfnamefont {E.}~\bibnamefont {Demler}},\
  }\bibfield  {title} {\enquote {\bibinfo {title} {Anomalous diffusion and
  griffiths effects near the many-body localization transition},}\ }\href
  {\doibase 10.1103/PhysRevLett.114.160401} {\bibfield  {journal} {\bibinfo
  {journal} {Phys. Rev. Lett.}\ }\textbf {\bibinfo {volume} {114}},\ \bibinfo
  {pages} {160401} (\bibinfo {year} {2015})}\BibitemShut {NoStop}%
\bibitem [{\citenamefont {Bar~Lev}\ \emph {et~al.}(2015)\citenamefont
  {Bar~Lev}, \citenamefont {Cohen},\ and\ \citenamefont {Reichman}}]{Lev15}%
  \BibitemOpen
  \bibfield  {author} {\bibinfo {author} {\bibfnamefont {Y.}~\bibnamefont
  {Bar~Lev}}, \bibinfo {author} {\bibfnamefont {G.}~\bibnamefont {Cohen}}, \
  and\ \bibinfo {author} {\bibfnamefont {D.~R.}\ \bibnamefont {Reichman}},\
  }\bibfield  {title} {\enquote {\bibinfo {title} {Absence of diffusion in an
  interacting system of spinless fermions on a one-dimensional disordered
  lattice},}\ }\href {\doibase 10.1103/PhysRevLett.114.100601} {\bibfield
  {journal} {\bibinfo  {journal} {Phys. Rev. Lett.}\ }\textbf {\bibinfo
  {volume} {114}},\ \bibinfo {pages} {100601} (\bibinfo {year}
  {2015})}\BibitemShut {NoStop}%
\bibitem [{\citenamefont {Torres-Herrera}\ and\ \citenamefont
  {Santos}(2015)}]{TorresHerrera15}%
  \BibitemOpen
  \bibfield  {author} {\bibinfo {author} {\bibfnamefont {E.~J.}\ \bibnamefont
  {Torres-Herrera}}\ and\ \bibinfo {author} {\bibfnamefont {Lea~F.}\
  \bibnamefont {Santos}},\ }\bibfield  {title} {\enquote {\bibinfo {title}
  {Dynamics at the many-body localization transition},}\ }\href {\doibase
  10.1103/PhysRevB.92.014208} {\bibfield  {journal} {\bibinfo  {journal} {Phys.
  Rev. B}\ }\textbf {\bibinfo {volume} {92}},\ \bibinfo {pages} {014208}
  (\bibinfo {year} {2015})}\BibitemShut {NoStop}%
\bibitem [{\citenamefont {Luitz}\ \emph {et~al.}(2016)\citenamefont {Luitz},
  \citenamefont {Laflorencie},\ and\ \citenamefont {Alet}}]{Luitz16}%
  \BibitemOpen
  \bibfield  {author} {\bibinfo {author} {\bibfnamefont {D.~J.}\ \bibnamefont
  {Luitz}}, \bibinfo {author} {\bibfnamefont {N.}~\bibnamefont {Laflorencie}},
  \ and\ \bibinfo {author} {\bibfnamefont {F.}~\bibnamefont {Alet}},\
  }\bibfield  {title} {\enquote {\bibinfo {title} {Extended slow dynamical
  regime close to the many-body localization transition},}\ }\href {\doibase
  10.1103/PhysRevB.93.060201} {\bibfield  {journal} {\bibinfo  {journal} {Phys.
  Rev. B}\ }\textbf {\bibinfo {volume} {93}},\ \bibinfo {pages} {060201}
  (\bibinfo {year} {2016})}\BibitemShut {NoStop}%
\bibitem [{\citenamefont {\ifmmode \check{Z}\else
  \v{Z}\fi{}nidari\ifmmode~\check{c}\else \v{c}\fi{}}\ \emph
  {et~al.}(2016)\citenamefont {\ifmmode \check{Z}\else
  \v{Z}\fi{}nidari\ifmmode~\check{c}\else \v{c}\fi{}}, \citenamefont
  {Scardicchio},\ and\ \citenamefont {Varma}}]{Znidaric16}%
  \BibitemOpen
  \bibfield  {author} {\bibinfo {author} {\bibfnamefont {M.}~\bibnamefont
  {\ifmmode \check{Z}\else \v{Z}\fi{}nidari\ifmmode~\check{c}\else
  \v{c}\fi{}}}, \bibinfo {author} {\bibfnamefont {A.}~\bibnamefont
  {Scardicchio}}, \ and\ \bibinfo {author} {\bibfnamefont {V.~K.}\ \bibnamefont
  {Varma}},\ }\bibfield  {title} {\enquote {\bibinfo {title} {Diffusive and
  subdiffusive spin transport in the ergodic phase of a many-body localizable
  system},}\ }\href {\doibase 10.1103/PhysRevLett.117.040601} {\bibfield
  {journal} {\bibinfo  {journal} {Phys. Rev. Lett.}\ }\textbf {\bibinfo
  {volume} {117}},\ \bibinfo {pages} {040601} (\bibinfo {year}
  {2016})}\BibitemShut {NoStop}%
\bibitem [{\citenamefont {Harris}(1974)}]{Harris74}%
  \BibitemOpen
  \bibfield  {author} {\bibinfo {author} {\bibfnamefont {A.~B.}\ \bibnamefont
  {Harris}},\ }\bibfield  {title} {\enquote {\bibinfo {title} {Effect of random
  defects on the critical behaviour of ising models},}\ }\href
  {http://stacks.iop.org/0022-3719/7/i=9/a=009} {\bibfield  {journal} {\bibinfo
   {journal} {J. Phys. Condens.}\ }\textbf {\bibinfo {volume} {7}},\ \bibinfo
  {pages} {1671} (\bibinfo {year} {1974})}\BibitemShut {NoStop}%
\bibitem [{\citenamefont {Chayes}\ \emph {et~al.}(1986)\citenamefont {Chayes},
  \citenamefont {Chayes}, \citenamefont {Fisher},\ and\ \citenamefont
  {Spencer}}]{Chayes86}%
  \BibitemOpen
  \bibfield  {author} {\bibinfo {author} {\bibfnamefont {J.~T.}\ \bibnamefont
  {Chayes}}, \bibinfo {author} {\bibfnamefont {L.}~\bibnamefont {Chayes}},
  \bibinfo {author} {\bibfnamefont {D.~S.}\ \bibnamefont {Fisher}}, \ and\
  \bibinfo {author} {\bibfnamefont {T.}~\bibnamefont {Spencer}},\ }\bibfield
  {title} {\enquote {\bibinfo {title} {Finite-size scaling and correlation
  lengths for disordered systems},}\ }\href {\doibase
  10.1103/PhysRevLett.57.2999} {\bibfield  {journal} {\bibinfo  {journal}
  {Phys. Rev. Lett.}\ }\textbf {\bibinfo {volume} {57}},\ \bibinfo {pages}
  {2999--3002} (\bibinfo {year} {1986})}\BibitemShut {NoStop}%
\bibitem [{\citenamefont {Chandran}\ \emph {et~al.}(2015)\citenamefont
  {Chandran}, \citenamefont {Laumann},\ and\ \citenamefont
  {Oganesyan}}]{Chandran15}%
  \BibitemOpen
  \bibfield  {author} {\bibinfo {author} {\bibfnamefont {A.}~\bibnamefont
  {Chandran}}, \bibinfo {author} {\bibfnamefont {C.~R.}\ \bibnamefont
  {Laumann}}, \ and\ \bibinfo {author} {\bibfnamefont {V.}~\bibnamefont
  {Oganesyan}},\ }\bibfield  {title} {\enquote {\bibinfo {title} {Finite size
  scaling bounds on many-body localized phase transitions},}\ }\href@noop {} {\
   (\bibinfo {year} {2015})},\ \Eprint {http://arxiv.org/abs/arXiv:1509.04285}
  {arXiv:1509.04285} \BibitemShut {NoStop}%
\bibitem [{\citenamefont {Khemani}\ \emph
  {et~al.}(2017{\natexlab{b}})\citenamefont {Khemani}, \citenamefont {Sheng},\
  and\ \citenamefont {Huse}}]{Khemani17}%
  \BibitemOpen
  \bibfield  {author} {\bibinfo {author} {\bibfnamefont {Vedika}\ \bibnamefont
  {Khemani}}, \bibinfo {author} {\bibfnamefont {D.~N.}\ \bibnamefont {Sheng}},
  \ and\ \bibinfo {author} {\bibfnamefont {David~A.}\ \bibnamefont {Huse}},\
  }\bibfield  {title} {\enquote {\bibinfo {title} {Two universality classes for
  the many-body localization transition},}\ }\href {\doibase
  10.1103/PhysRevLett.119.075702} {\bibfield  {journal} {\bibinfo  {journal}
  {Phys. Rev. Lett.}\ }\textbf {\bibinfo {volume} {119}},\ \bibinfo {pages}
  {075702} (\bibinfo {year} {2017}{\natexlab{b}})}\BibitemShut {NoStop}%
\bibitem [{\citenamefont {Gopalakrishnan}\ \emph {et~al.}(2016)\citenamefont
  {Gopalakrishnan}, \citenamefont {Agarwal}, \citenamefont {Demler},
  \citenamefont {Huse},\ and\ \citenamefont {Knap}}]{Gopalakrishnan16}%
  \BibitemOpen
  \bibfield  {author} {\bibinfo {author} {\bibfnamefont {S.}~\bibnamefont
  {Gopalakrishnan}}, \bibinfo {author} {\bibfnamefont {K.}~\bibnamefont
  {Agarwal}}, \bibinfo {author} {\bibfnamefont {E.~A.}\ \bibnamefont {Demler}},
  \bibinfo {author} {\bibfnamefont {D.~A.}\ \bibnamefont {Huse}}, \ and\
  \bibinfo {author} {\bibfnamefont {M.}~\bibnamefont {Knap}},\ }\bibfield
  {title} {\enquote {\bibinfo {title} {Griffiths effects and slow dynamics in
  nearly many-body localized systems},}\ }\href {\doibase
  10.1103/PhysRevB.93.134206} {\bibfield  {journal} {\bibinfo  {journal} {Phys.
  Rev. B}\ }\textbf {\bibinfo {volume} {93}},\ \bibinfo {pages} {134206}
  (\bibinfo {year} {2016})}\BibitemShut {NoStop}%
\bibitem [{\citenamefont {Lev}\ \emph {et~al.}(2017)\citenamefont {Lev},
  \citenamefont {Kennes}, \citenamefont {Klöckner}, \citenamefont {Reichman},\
  and\ \citenamefont {Karrasch}}]{BarLev17}%
  \BibitemOpen
  \bibfield  {author} {\bibinfo {author} {\bibfnamefont {Y.~Bar}\ \bibnamefont
  {Lev}}, \bibinfo {author} {\bibfnamefont {D.~M.}\ \bibnamefont {Kennes}},
  \bibinfo {author} {\bibfnamefont {C.}~\bibnamefont {Klöckner}}, \bibinfo
  {author} {\bibfnamefont {D.~R.}\ \bibnamefont {Reichman}}, \ and\ \bibinfo
  {author} {\bibfnamefont {C.}~\bibnamefont {Karrasch}},\ }\bibfield  {title}
  {\enquote {\bibinfo {title} {Transport in quasiperiodic interacting systems:
  From superdiffusion to subdiffusion},}\ }\href
  {http://stacks.iop.org/0295-5075/119/i=3/a=37003} {\bibfield  {journal}
  {\bibinfo  {journal} {Europhys. Lett.}\ }\textbf {\bibinfo {volume} {119}},\
  \bibinfo {pages} {37003} (\bibinfo {year} {2017})}\BibitemShut {NoStop}%
\bibitem [{\citenamefont {Aubry}\ and\ \citenamefont
  {Andr{\'e}}(1980)}]{Aubry80}%
  \BibitemOpen
  \bibfield  {author} {\bibinfo {author} {\bibfnamefont {S.}~\bibnamefont
  {Aubry}}\ and\ \bibinfo {author} {\bibfnamefont {G.}~\bibnamefont
  {Andr{\'e}}},\ }\bibfield  {title} {\enquote {\bibinfo {title} {Analyticity
  breaking and anderson localization in incommensurate lattices},}\ }\href@noop
  {} {\bibfield  {journal} {\bibinfo  {journal} {Ann. Israel Phys. Soc.}\
  }\textbf {\bibinfo {volume} {3}},\ \bibinfo {pages} {18} (\bibinfo {year}
  {1980})}\BibitemShut {NoStop}%
\bibitem [{SOM()}]{SOMs}%
  \BibitemOpen
  \href@noop {} {}\bibinfo {note} {See Supplementary Material for details,
  including Refs.~\cite{Regal03,SLEPC}}\BibitemShut {NoStop}%
\bibitem [{\citenamefont {Trotzky}\ \emph {et~al.}(2012)\citenamefont
  {Trotzky}, \citenamefont {Chen}, \citenamefont {Flesch}, \citenamefont
  {McCulloch}, \citenamefont {Schollw{\"o}ck}, \citenamefont {Eisert},\ and\
  \citenamefont {Bloch}}]{Trotzky12}%
  \BibitemOpen
  \bibfield  {author} {\bibinfo {author} {\bibfnamefont {S.}~\bibnamefont
  {Trotzky}}, \bibinfo {author} {\bibfnamefont {Y-A.}\ \bibnamefont {Chen}},
  \bibinfo {author} {\bibfnamefont {A.}~\bibnamefont {Flesch}}, \bibinfo
  {author} {\bibfnamefont {I.~P.}\ \bibnamefont {McCulloch}}, \bibinfo {author}
  {\bibfnamefont {U.}~\bibnamefont {Schollw{\"o}ck}}, \bibinfo {author}
  {\bibfnamefont {J.}~\bibnamefont {Eisert}}, \ and\ \bibinfo {author}
  {\bibfnamefont {I.}~\bibnamefont {Bloch}},\ }\bibfield  {title} {\enquote
  {\bibinfo {title} {Probing the relaxation towards equilibrium in an isolated
  strongly correlated one-dimensional bose gas},}\ }\href
  {http://dx.doi.org/10.1038/nphys2232} {\bibfield  {journal} {\bibinfo
  {journal} {Nat. Phys.}\ }\textbf {\bibinfo {volume} {8}},\ \bibinfo {pages}
  {325} (\bibinfo {year} {2012})}\BibitemShut {NoStop}%
\bibitem [{\citenamefont {Lux}\ \emph {et~al.}(2014)\citenamefont {Lux},
  \citenamefont {M{\"u}ller}, \citenamefont {Mitra},\ and\ \citenamefont
  {Rosch}}]{Rosch14}%
  \BibitemOpen
  \bibfield  {author} {\bibinfo {author} {\bibfnamefont {J.}~\bibnamefont
  {Lux}}, \bibinfo {author} {\bibfnamefont {J.}~\bibnamefont {M{\"u}ller}},
  \bibinfo {author} {\bibfnamefont {A.}~\bibnamefont {Mitra}}, \ and\ \bibinfo
  {author} {\bibfnamefont {A.}~\bibnamefont {Rosch}},\ }\bibfield  {title}
  {\enquote {\bibinfo {title} {Hydrodynamic long-time tails after a quantum
  quench},}\ }\href {\doibase 10.1103/PhysRevA.89.053608} {\bibfield  {journal}
  {\bibinfo  {journal} {Phys. Rev. A}\ }\textbf {\bibinfo {volume} {89}},\
  \bibinfo {pages} {053608} (\bibinfo {year} {2014})}\BibitemShut {NoStop}%
\bibitem [{\citenamefont {Bordia}\ \emph {et~al.}(2016)\citenamefont {Bordia},
  \citenamefont {L\"uschen}, \citenamefont {Hodgman}, \citenamefont
  {Schreiber}, \citenamefont {Bloch},\ and\ \citenamefont
  {Schneider}}]{Bordia16}%
  \BibitemOpen
  \bibfield  {author} {\bibinfo {author} {\bibfnamefont {P.}~\bibnamefont
  {Bordia}}, \bibinfo {author} {\bibfnamefont {H.~P.}\ \bibnamefont
  {L\"uschen}}, \bibinfo {author} {\bibfnamefont {S.~S.}\ \bibnamefont
  {Hodgman}}, \bibinfo {author} {\bibfnamefont {M.}~\bibnamefont {Schreiber}},
  \bibinfo {author} {\bibfnamefont {I.}~\bibnamefont {Bloch}}, \ and\ \bibinfo
  {author} {\bibfnamefont {U.}~\bibnamefont {Schneider}},\ }\bibfield  {title}
  {\enquote {\bibinfo {title} {Coupling identical one-dimensional many-body
  localized systems},}\ }\href {\doibase 10.1103/PhysRevLett.116.140401}
  {\bibfield  {journal} {\bibinfo  {journal} {Phys. Rev. Lett.}\ }\textbf
  {\bibinfo {volume} {116}},\ \bibinfo {pages} {140401} (\bibinfo {year}
  {2016})}\BibitemShut {NoStop}%
\bibitem [{\citenamefont {L{\"u}schen}\ \emph {et~al.}(2017)\citenamefont
  {L{\"u}schen}, \citenamefont {Bordia}, \citenamefont {Hodgman}, \citenamefont
  {Schreiber}, \citenamefont {Sarkar}, \citenamefont {Daley}, \citenamefont
  {Fischer}, \citenamefont {Altman}, \citenamefont {Bloch},\ and\ \citenamefont
  {Schneider}}]{Lueschen16}%
  \BibitemOpen
  \bibfield  {author} {\bibinfo {author} {\bibfnamefont {H.~P.}\ \bibnamefont
  {L{\"u}schen}}, \bibinfo {author} {\bibfnamefont {P.}~\bibnamefont {Bordia}},
  \bibinfo {author} {\bibfnamefont {S.~S.}\ \bibnamefont {Hodgman}}, \bibinfo
  {author} {\bibfnamefont {M.}~\bibnamefont {Schreiber}}, \bibinfo {author}
  {\bibfnamefont {S.}~\bibnamefont {Sarkar}}, \bibinfo {author} {\bibfnamefont
  {A.~J.}\ \bibnamefont {Daley}}, \bibinfo {author} {\bibfnamefont {M.~H.}\
  \bibnamefont {Fischer}}, \bibinfo {author} {\bibfnamefont {E.}~\bibnamefont
  {Altman}}, \bibinfo {author} {\bibfnamefont {I.}~\bibnamefont {Bloch}}, \
  and\ \bibinfo {author} {\bibfnamefont {U.}~\bibnamefont {Schneider}},\
  }\bibfield  {title} {\enquote {\bibinfo {title} {Signatures of many-body
  localization in a controlled open quantum system},}\ }\href {\doibase
  10.1103/PhysRevX.7.011034} {\bibfield  {journal} {\bibinfo  {journal} {Phys.
  Rev. X}\ }\textbf {\bibinfo {volume} {7}},\ \bibinfo {pages} {011034}
  (\bibinfo {year} {2017})}\BibitemShut {NoStop}%
\bibitem [{\citenamefont {Lee}\ \emph {et~al.}(2017)\citenamefont {Lee},
  \citenamefont {Look}, \citenamefont {Lim},\ and\ \citenamefont
  {Sheng}}]{Lee17}%
  \BibitemOpen
  \bibfield  {author} {\bibinfo {author} {\bibfnamefont {Mac}\ \bibnamefont
  {Lee}}, \bibinfo {author} {\bibfnamefont {Thomas~R.}\ \bibnamefont {Look}},
  \bibinfo {author} {\bibfnamefont {S.~P.}\ \bibnamefont {Lim}}, \ and\
  \bibinfo {author} {\bibfnamefont {D.~N.}\ \bibnamefont {Sheng}},\ }\bibfield
  {title} {\enquote {\bibinfo {title} {Many-body localization in spin chain
  systems with quasiperiodic fields},}\ }\href {\doibase
  10.1103/PhysRevB.96.075146} {\bibfield  {journal} {\bibinfo  {journal} {Phys.
  Rev. B}\ }\textbf {\bibinfo {volume} {96}},\ \bibinfo {pages} {075146}
  (\bibinfo {year} {2017})}\BibitemShut {NoStop}%
\bibitem [{\citenamefont {Mierzejewski}\ \emph {et~al.}(2016)\citenamefont
  {Mierzejewski}, \citenamefont {Herbrych},\ and\ \citenamefont
  {Prelov\v{s}ek}}]{Mierzejewski16}%
  \BibitemOpen
  \bibfield  {author} {\bibinfo {author} {\bibfnamefont {M.}~\bibnamefont
  {Mierzejewski}}, \bibinfo {author} {\bibfnamefont {J.}~\bibnamefont
  {Herbrych}}, \ and\ \bibinfo {author} {\bibfnamefont {P.}~\bibnamefont
  {Prelov\v{s}ek}},\ }\bibfield  {title} {\enquote {\bibinfo {title} {Universal
  dynamics of density correlations at the transition to the many-body localized
  state},}\ }\href {\doibase 10.1103/PhysRevB.94.224207} {\bibfield  {journal}
  {\bibinfo  {journal} {Phys. Rev. B}\ }\textbf {\bibinfo {volume} {94}},\
  \bibinfo {pages} {224207} (\bibinfo {year} {2016})}\BibitemShut {NoStop}%
\bibitem [{\citenamefont {Bordia}\ \emph {et~al.}(2017)\citenamefont {Bordia},
  \citenamefont {L{\"u}schen}, \citenamefont {Schneider}, \citenamefont
  {Knap},\ and\ \citenamefont {Bloch}}]{Bordia16_2}%
  \BibitemOpen
  \bibfield  {author} {\bibinfo {author} {\bibfnamefont {P.}~\bibnamefont
  {Bordia}}, \bibinfo {author} {\bibfnamefont {H.}~\bibnamefont {L{\"u}schen}},
  \bibinfo {author} {\bibfnamefont {U.}~\bibnamefont {Schneider}}, \bibinfo
  {author} {\bibfnamefont {M.}~\bibnamefont {Knap}}, \ and\ \bibinfo {author}
  {\bibfnamefont {I.}~\bibnamefont {Bloch}},\ }\bibfield  {title} {\enquote
  {\bibinfo {title} {Periodically driving a many-body localized quantum
  system},}\ }\href {\doibase http://dx.doi.org/10.1038/nphys4020} {\bibfield
  {journal} {\bibinfo  {journal} {Nat Phys}\ }\textbf {\bibinfo {volume}
  {13}},\ \bibinfo {pages} {460--464} (\bibinfo {year} {2017})}\BibitemShut
  {NoStop}%
\bibitem [{\citenamefont {Vasseur}\ \emph {et~al.}(2015)\citenamefont
  {Vasseur}, \citenamefont {Potter},\ and\ \citenamefont
  {Parameswaran}}]{Vasseur15}%
  \BibitemOpen
  \bibfield  {author} {\bibinfo {author} {\bibfnamefont {R.}~\bibnamefont
  {Vasseur}}, \bibinfo {author} {\bibfnamefont {A.~C.}\ \bibnamefont {Potter}},
  \ and\ \bibinfo {author} {\bibfnamefont {S.~A.}\ \bibnamefont
  {Parameswaran}},\ }\bibfield  {title} {\enquote {\bibinfo {title} {Quantum
  criticality of hot random spin chains},}\ }\href {\doibase
  10.1103/PhysRevLett.114.217201} {\bibfield  {journal} {\bibinfo  {journal}
  {Phys. Rev. Lett.}\ }\textbf {\bibinfo {volume} {114}},\ \bibinfo {pages}
  {217201} (\bibinfo {year} {2015})}\BibitemShut {NoStop}%
\bibitem [{\citenamefont {Prelov\ifmmode~\check{s}\else \v{s}\fi{}ek}\ \emph
  {et~al.}(2016)\citenamefont {Prelov\ifmmode~\check{s}\else \v{s}\fi{}ek},
  \citenamefont {Bari\ifmmode \check{s}\else \v{s}\fi{}i\ifmmode~\acute{c}\else
  \'{c}\fi{}},\ and\ \citenamefont {\ifmmode \check{Z}\else
  \v{Z}\fi{}nidari\ifmmode~\check{c}\else \v{c}\fi{}}}]{Prelovsek16}%
  \BibitemOpen
  \bibfield  {author} {\bibinfo {author} {\bibfnamefont {P.}~\bibnamefont
  {Prelov\ifmmode~\check{s}\else \v{s}\fi{}ek}}, \bibinfo {author}
  {\bibfnamefont {O.~S.}\ \bibnamefont {Bari\ifmmode \check{s}\else
  \v{s}\fi{}i\ifmmode~\acute{c}\else \'{c}\fi{}}}, \ and\ \bibinfo {author}
  {\bibfnamefont {M.}~\bibnamefont {\ifmmode \check{Z}\else
  \v{Z}\fi{}nidari\ifmmode~\check{c}\else \v{c}\fi{}}},\ }\bibfield  {title}
  {\enquote {\bibinfo {title} {Absence of full many-body localization in the
  disordered {H}ubbard chain},}\ }\href {\doibase 10.1103/PhysRevB.94.241104}
  {\bibfield  {journal} {\bibinfo  {journal} {Phys. Rev. B}\ }\textbf {\bibinfo
  {volume} {94}},\ \bibinfo {pages} {241104} (\bibinfo {year}
  {2016})}\BibitemShut {NoStop}%
\bibitem [{\citenamefont {Protopopov}\ \emph {et~al.}(2017)\citenamefont
  {Protopopov}, \citenamefont {Ho},\ and\ \citenamefont
  {Abanin}}]{Protopopov16}%
  \BibitemOpen
  \bibfield  {author} {\bibinfo {author} {\bibfnamefont {Ivan~V.}\ \bibnamefont
  {Protopopov}}, \bibinfo {author} {\bibfnamefont {Wen~Wei}\ \bibnamefont
  {Ho}}, \ and\ \bibinfo {author} {\bibfnamefont {Dmitry~A.}\ \bibnamefont
  {Abanin}},\ }\bibfield  {title} {\enquote {\bibinfo {title} {Effect of
  {SU(2)} symmetry on many-body localization and thermalization},}\ }\href
  {\doibase 10.1103/PhysRevB.96.041122} {\bibfield  {journal} {\bibinfo
  {journal} {Phys. Rev. B}\ }\textbf {\bibinfo {volume} {96}},\ \bibinfo
  {pages} {041122} (\bibinfo {year} {2017})}\BibitemShut {NoStop}%
\bibitem [{\citenamefont {Lucioni}\ \emph {et~al.}(2011)\citenamefont
  {Lucioni}, \citenamefont {Deissler}, \citenamefont {Tanzi}, \citenamefont
  {Roati}, \citenamefont {Zaccanti}, \citenamefont {Modugno}, \citenamefont
  {Larcher}, \citenamefont {Dalfovo}, \citenamefont {Inguscio},\ and\
  \citenamefont {Modugno}}]{Lucioni11}%
  \BibitemOpen
  \bibfield  {author} {\bibinfo {author} {\bibfnamefont {E.}~\bibnamefont
  {Lucioni}}, \bibinfo {author} {\bibfnamefont {B.}~\bibnamefont {Deissler}},
  \bibinfo {author} {\bibfnamefont {L.}~\bibnamefont {Tanzi}}, \bibinfo
  {author} {\bibfnamefont {G.}~\bibnamefont {Roati}}, \bibinfo {author}
  {\bibfnamefont {M.}~\bibnamefont {Zaccanti}}, \bibinfo {author}
  {\bibfnamefont {M.}~\bibnamefont {Modugno}}, \bibinfo {author} {\bibfnamefont
  {M.}~\bibnamefont {Larcher}}, \bibinfo {author} {\bibfnamefont
  {F.}~\bibnamefont {Dalfovo}}, \bibinfo {author} {\bibfnamefont
  {M.}~\bibnamefont {Inguscio}}, \ and\ \bibinfo {author} {\bibfnamefont
  {G.}~\bibnamefont {Modugno}},\ }\bibfield  {title} {\enquote {\bibinfo
  {title} {Observation of subdiffusion in a disordered interacting system},}\
  }\href {\doibase 10.1103/PhysRevLett.106.230403} {\bibfield  {journal}
  {\bibinfo  {journal} {Phys. Rev. Lett.}\ }\textbf {\bibinfo {volume} {106}},\
  \bibinfo {pages} {230403} (\bibinfo {year} {2011})}\BibitemShut {NoStop}%
\bibitem [{\citenamefont {Regal}\ \emph {et~al.}(2003)\citenamefont {Regal},
  \citenamefont {Ticknor}, \citenamefont {Bohn},\ and\ \citenamefont
  {Jin}}]{Regal03}%
  \BibitemOpen
  \bibfield  {author} {\bibinfo {author} {\bibfnamefont {C.~A.}\ \bibnamefont
  {Regal}}, \bibinfo {author} {\bibfnamefont {C.}~\bibnamefont {Ticknor}},
  \bibinfo {author} {\bibfnamefont {J.~L.}\ \bibnamefont {Bohn}}, \ and\
  \bibinfo {author} {\bibfnamefont {D.~S.}\ \bibnamefont {Jin}},\ }\bibfield
  {title} {\enquote {\bibinfo {title} {Creation of ultracold molecules from a
  fermi gas of atoms},}\ }\href {http://dx.doi.org/10.1038/nature01738}
  {\bibfield  {journal} {\bibinfo  {journal} {Nature}\ }\textbf {\bibinfo
  {volume} {424}},\ \bibinfo {pages} {47} (\bibinfo {year} {2003})}\BibitemShut
  {NoStop}%
\bibitem [{\citenamefont {Hernandez}\ \emph {et~al.}(2005)\citenamefont
  {Hernandez}, \citenamefont {Roman},\ and\ \citenamefont {Vidal}}]{SLEPC}%
  \BibitemOpen
  \bibfield  {author} {\bibinfo {author} {\bibfnamefont {V.}~\bibnamefont
  {Hernandez}}, \bibinfo {author} {\bibfnamefont {J.~E.}\ \bibnamefont
  {Roman}}, \ and\ \bibinfo {author} {\bibfnamefont {V.}~\bibnamefont
  {Vidal}},\ }\bibfield  {title} {\enquote {\bibinfo {title} {{SLEPc}: A
  scalable and flexible toolkit for the solution of eigenvalue problems},}\
  }\href {\doibase 10.1145/1089014.1089019} {\bibfield  {journal} {\bibinfo
  {journal} {{ACM} Trans. Math. Software}\ }\textbf {\bibinfo {volume} {31}},\
  \bibinfo {pages} {351--362} (\bibinfo {year} {2005})}\BibitemShut {NoStop}%
\end{thebibliography}%

\newpage

\appendix

\setcounter{figure}{0}
\setcounter{equation}{0}

\renewcommand{\thepage}{S\arabic{page}} 
\renewcommand{\thesection}{S\arabic{section}}  
\renewcommand{\thetable}{S\arabic{table}}  
\renewcommand{\thefigure}{S\arabic{figure}}

\section{Supplementary Material}

\subsection{Comparison to the noninteracting system}

\begin{figure}
	\centering
	\includegraphics[width=84mm]{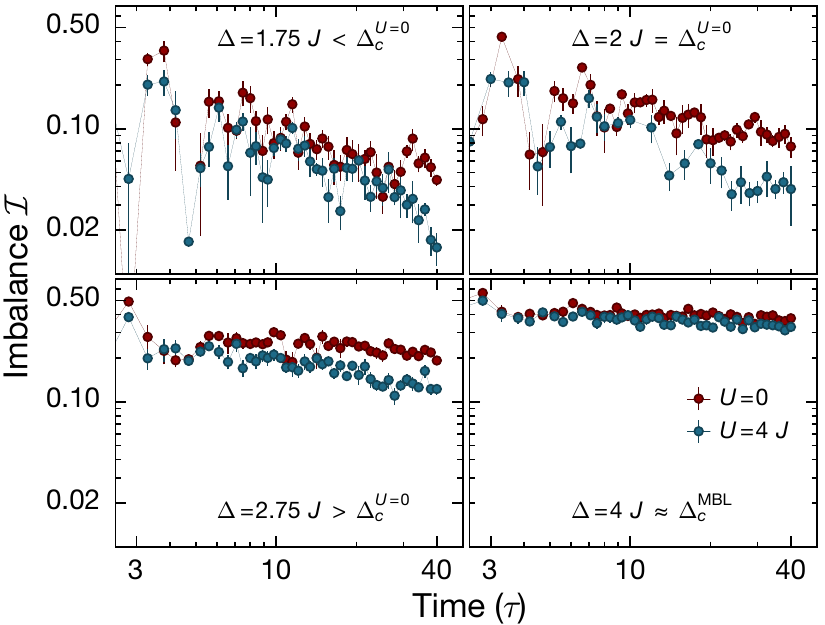}
	\caption{\textbf{Interacting vs.\ noninteracting time traces:} Comparison between the decay of an initially prepared charge-density wave in the absence and presence of interactions for various disorder strengths $\Delta$. Each point is the average over 6 disorder phases $\phi$, with error bars indicating the uncertainty of the mean. During the first three tunneling times (not shown), the imbalance quickly decays from its initial value close to one before the dynamics crosses over into the shown, much slower regime.}
	\label{SOMs_traces_fig}
\end{figure}

We perform a detailed comparison of the interacting and noninteracting slow relaxation dynamics. Fig.~\ref{SOMs_traces_fig} compares the imbalance time traces for disorder strengths below and close to the noninteracting transition point ($\Delta_c^{U=0} = 2\, J$), as well as below and close to the estimated interacting transition. The data omits the fast initial decay of the imbalance, and shows a slow, further decay in the time window spanned by the 'short' ($10 \, \tau$) and 'long' ($40 \, \tau$) observation times of Fig.~1.

Below the noninteracting transition (at $\Delta = 1.75 \, J$) the interacting and noninteracting imbalance behave very similar and quickly decay to zero. This indicates that the dynamics is dominated by the spreading of single particles. This picture changes upon approaching the single-particle localization transition (at $\Delta_c^{U=0} = 2 \, J$), where the noninteracting system already shows a significantly slower decay. We attribute this to the fact that the noninteracting system is already localized, and the observed imbalance decay is part of the initial relaxation during which the particles spread within their single-particle localization length. Further beyond the single-particle transition (at $\Delta = 2.75 \, J$), where the single-particle localization length is small, the noninteracting system is completely frozen and its imbalance remains stationary. Yet, the interacting system continues to show a slow decay, as is discussed in the main text (see Fig.~2). The decay in the interacting system only stops at even larger disorder strengths ($\Delta \gtrsim 4 \, J$) when the system becomes many-body localized. In the MBL regime, the interacting imbalance still lies slightly below the noninteracting values, likely indicating an increased localization length. In addition, however, the effects of external bath couplings, which are larger in the presence of interactions~\cite{Bordia16,Lueschen16}, also lower the interacting imbalance. Furthermore, intermediate, possibly logarithmic decays towards a finite imbalance, which have been found to occur in the MBL phase~\cite{Mierzejewski16}, could contribute.

\begin{figure}
	\centering
	\includegraphics[width=84mm]{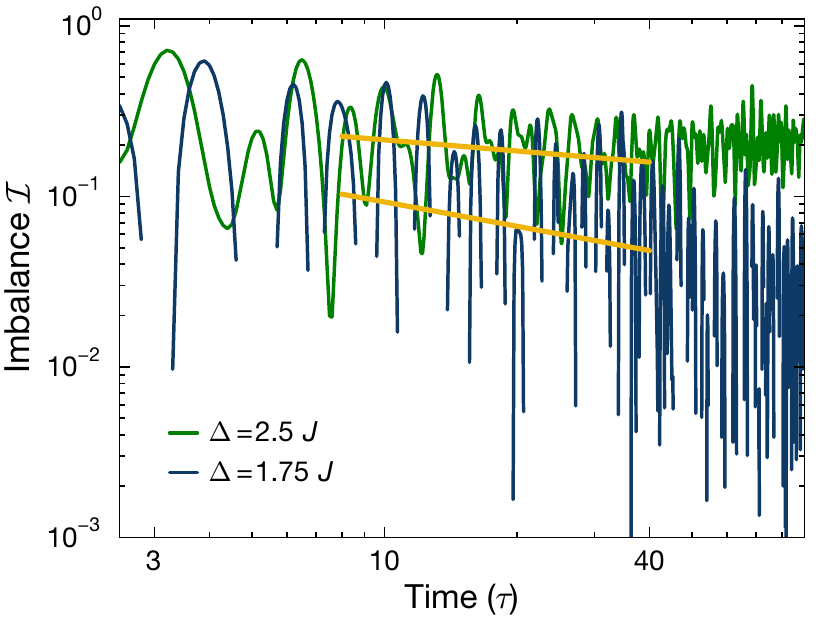}
	\caption{\textbf{Simulation of noninteracting time traces:} Imbalance decay of an initial CDW of noninteracting particles for $\Delta = 1.75 \, J$ and $\Delta = 2.5 \, J$ on a system with $200$ sites, averaged over $1000$ phases $\phi$. The fast initial decay of the imbalance from its starting value is omitted. We observe much stronger oscillations than in the experiment, but reproduce the general trend. This is illustrated by the yellow lines, which show fits to the \textit{experimental} data on the time scales of the experiment.}
	\label{SOMs_nonint_ED_fig}
\end{figure}

\begin{figure}
	\centering
	\includegraphics[width=84mm]{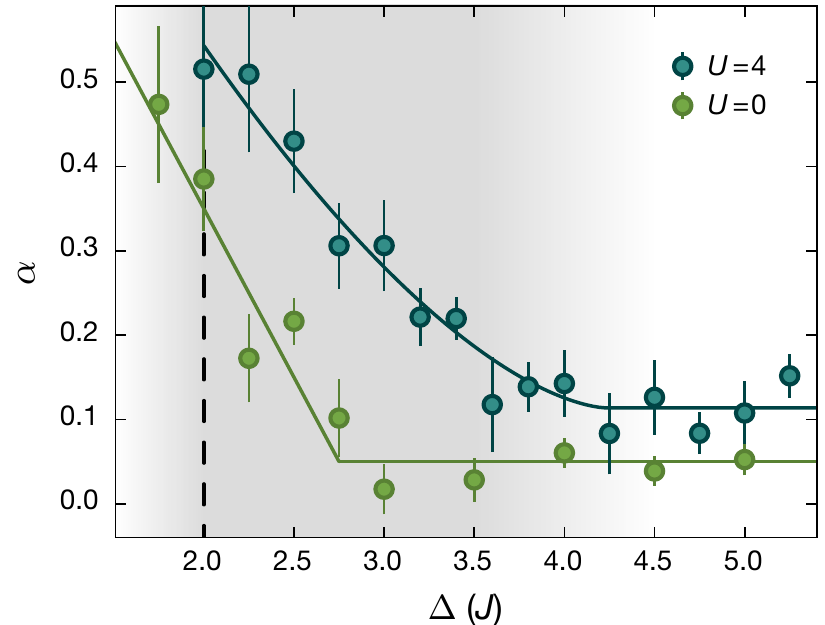}
	\caption{\textbf{Interacting vs. noninteracting exponents:} Exponents of a power-law fit to the interacting and noninteracting time traces between $8 - 40 \, \tau$. Errorbars denote the covariance error of the fit. Solid lines are guides-to-the-eye.}
	\label{SOMs_naive_exponents_fig}
\end{figure}

We note that the above comparison is difficult to perform with theoretical data, as the noninteracting system exhibits strong oscillations (see Fig.~\ref{SOMs_nonint_ED_fig}). In the experiment these oscillations quickly dephase due to averaging over many 1D tubes with slightly different tunneling rates and disorder strengths. While the functional form of the averaged noninteracting dynamics is not clear, on the accessed time scales the tube-averaged decays can be well fitted with power laws, enabling a direct comparison of the interacting and noninteracting exponents $\alpha$ shown in Fig.~\ref{SOMs_naive_exponents_fig}. Both the interacting and the noninteracting exponents decrease with increasing disorder strength around the noninteracting localization transition until eventually both saturate at their respective offset values. The offset value of the noninteracting system is slightly smaller than the offset of the interacting system, which is expected as external couplings have a stronger influence on the interacting system~\cite{Bordia16,Lueschen16}. We find the saturation of the noninteracting exponents to occur at a slightly larger disorder strength than $\Delta_c^{U=0} = 2 \, J$. We attribute this to the slow spreading of single particles to their localization length, which is larger than one lattice site close to the transition. Full freezing of the particles only happens when the localization length drops below one lattice site. In the interacting system, we believe that similar effects impact our analysis of the critical point much less, as such intermediate relaxations above the MBL critical point are expected to be logarithmically slow~\cite{Prelovsek16}.

\subsection{Interaction dependence of the power-law exponents}

\begin{figure}
	\centering
	\includegraphics[width=84mm]{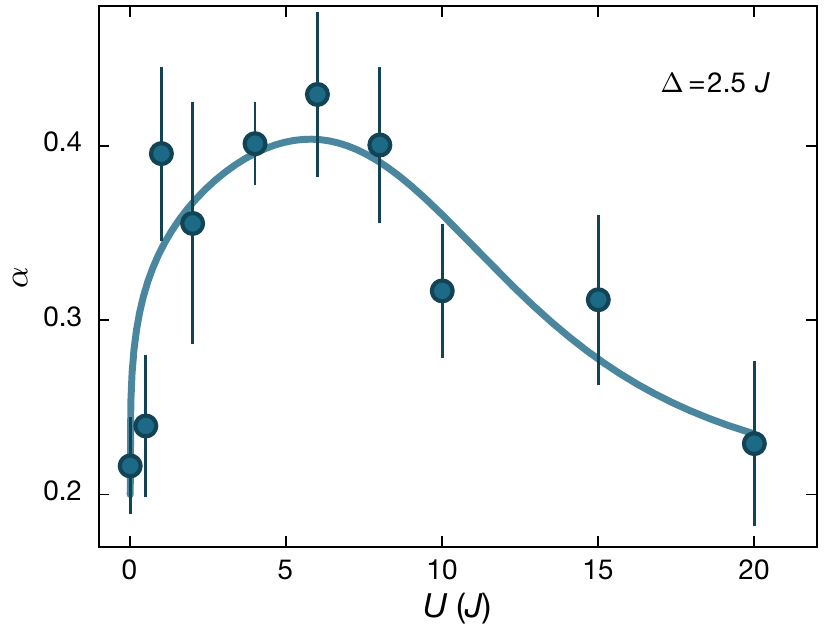}
	\caption{\textbf{Power-law exponents versus interaction strength:} Fitted experimental exponents as a function of the interaction strength $U$ at a disorder strength of $\Delta = 2.5 \, J$.}
	\label{SOMs_along_U_fig}
\end{figure}

In this work, we performed the analysis of the interacting system at a fixed interaction strength of $U = 4 \, J$. The value was chosen, since here we expect the strongest interaction shift of the MBL transition compared to the noninteracting localization transition~\cite{Schreiber15}. Fig.~\ref{SOMs_along_U_fig} shows the extracted power-law exponents as a function of interaction strength at $\Delta = 2.5 \, J$. We find that the exponent shows a behavior reminiscent of the interaction dependence observed in Ref.~\cite{Schreiber15}: We find a maximum interaction effect around $U=5 \, J$ and values approaching the noninteracting limit in the regime of hard core interactions. This is expected due to an exact mathematical mapping from hardcore fermions to free fermions in the absence of doubly occupied lattice sites~\cite{Schreiber15}.

\subsection{Details of the experimental setup}
We start our experiments by cooling an equal mixture of the two lowest lying hyperfine states (denoted as $\ket{\uparrow}$,$\ket{\downarrow}$) of $^{40}$K atoms to a temperature of $0.15\,T/T_\mathrm{F}$ in a dipole trap, where $T_\mathrm{F}$ is the Fermi temperature. The cold gas is then loaded into a three dimensional optical lattice, where the formation of doublons is suppressed to below $\approx 8\%$ by strong repulsive interactions. 

The optical lattice consists of two deep $\lambda_{\mathrm {\perp}} \approx 738\,$nm lattices along the orthogonal directions at a strength of $40\,\mathrm{E}^{\perp}_{\mathrm{r}}$, creating an array of one-dimensional tubes. Here $E_r^i=\mathrm{h}^2 / (2 m \lambda_i^2)$ denotes the recoil energy, where $\lambda_i$ is the respective lattice's wavelength and $m$ is the mass of $^{40}$K. In the tubes, we employ a primary $\lambda_p \approx 532\,$nm lattice, which is superimposed by a weaker, incommensurate $\lambda_d \approx 738.2\,$nm disorder lattice, to implement the Aubry-Andr{\'e} model~\cite{Aubry80} with on-site interactions~\cite{Iyer13}.

\begin{equation}
\begin{split}
\hat{H} = &-J\sum_{j,\sigma}(\hat{c}^{\dagger}_{j+1,\sigma}\hat{c}_{j,\sigma} + \text{h.c.}) \\& + \Delta\sum_{j,\sigma} \cos (2\pi\beta j+\phi)\hat{n}_{j,\sigma} +U\sum_{j} \hat{n}_{j,\uparrow}\hat{n}_{j,\downarrow},
\label{AA_hamiltonian}
\end{split}
\end{equation}

\noindent
The operators $\hat{c}^{\dagger}_{j,\sigma}$ and $\hat{c}_{j,\sigma}$ are the creation and annihilation operators for spin $\sigma \in \{ \ket{ \uparrow }, \ket{ \downarrow } \}$ on site $j$, the respective number operators are given by $\hat{n}_{j,\uparrow}$ and $\hat{n}_{j,\downarrow}$.
$J \approx \mathrm{h} \times 500\,$Hz denotes the tunneling rate in the primary lattice. The strength of the correlated disorder $\Delta$ and its phase $\phi$ is controlled by the depth and relative phase of the disorder lattice. The incommensurable ratio is given by $\beta = \lambda_d/\lambda_p$. Additionally, we achieve independent control of the interaction strength $U$ via a Feshbach resonance centered around 202.1\,G~\cite{Regal03}. We note that there are small deviations between our setup and an ideal Aubry-Andre{\'e} model as, e.g.\ the disorder lattice not only modulates the on-site energies, but also the hopping rates by $\lesssim 10 \%$. Details of these deviations can be found in Ref.~\cite{Schreiber15}.

We create the initial charge-density wave using an additional superlattice with respect to the primary lattice with $\lambda_l \approx 1064\,$nm, by following the same loading procedure as in Ref.~\cite{Schreiber15}. After the evolution time, the superlattice is employed again to extract the imbalance $\mathcal{I}$ via a band mapping technique~\cite{Trotzky12}.
\\

\subsection{Power-law versus exponential fits}

\begin{figure}
	\centering
	\includegraphics[width=84mm]{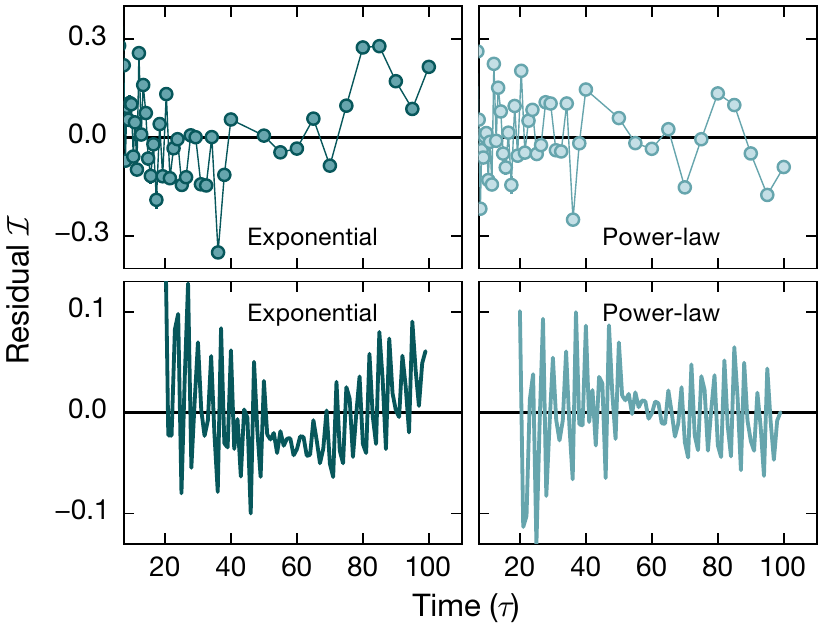}
	\caption{\textbf{Residuals of exponential and power-law fits:} Residuals of the Imbalance at $\Delta = 2.5 \, J$ calculated as $\log(\mathcal{I})-\log(\mathcal{I}_\mathrm{fit})$ for experiment and theory for exponential and power-law fits as in Fig.~2b. Theory data (lines) is only shown from $20 \, \tau$, as earlier oscillations strongly affect the fit.}
	\label{SOMs_residuals_fig}
\end{figure}

Fig.~\ref{SOMs_residuals_fig} shows the residuals of the exponential and power-law fits to the imbalance decay at $\Delta = 2.5 \, J$ shown in Fig.~2b. In the experimental data the residuals of the power-law fit scatter symmetrically around zero. The exponential residuals are below zero in-between $20-40 \, \tau$ and above zero beyond $80 \, \tau$, indicating that a power-law fit captures the data slightly better. A similar trend can be seen in the theory, where only the exponential residuals show a systematic trend. 
\\

\subsection{Analysis of the dynamics using exponential fits}

\begin{figure}
	\centering
	\includegraphics[width=84mm]{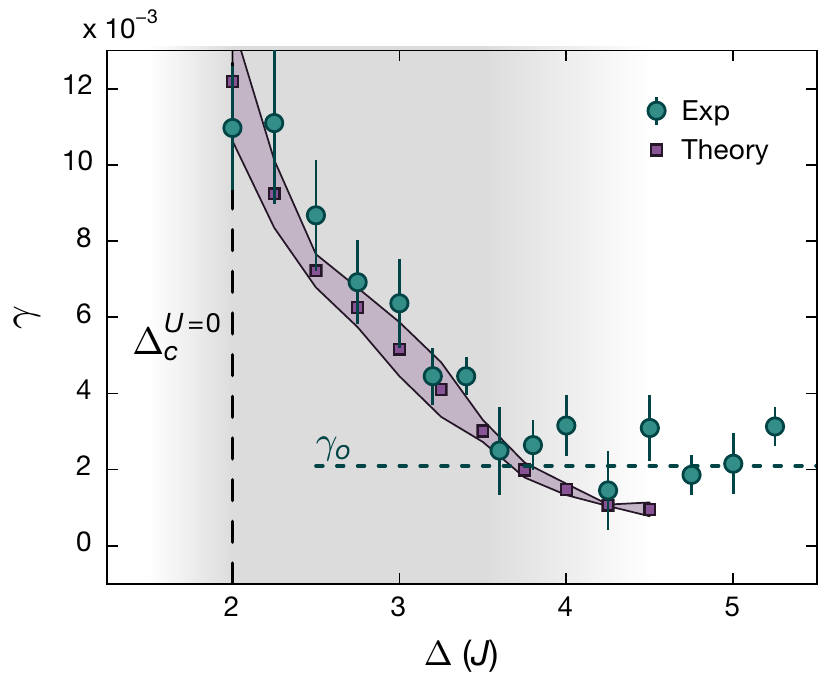}
	\caption{\textbf{Exponential fit analysis of the slow dynamics:} Analysis of the slow dynamics equivalent to Fig.~3 in the main text, but using exponential fits. Plotted is the decay rate $\gamma$. Errorbars indicate uncertainty of the fit to the experimental data, the purple shaded region an estimate of the finite size error of the theoretical simulations. For the largest disorder strengths, systematic finite time and size errors do not allow an accurate fit of the simulation data and the uncertainty is likely underestimated. The gray shading marks the regime of slow dynamics from Fig.~1 in the main text. As in the analysis with power-law fits, we observe a saturation of the experimental data to an offset value of $1/\tau_0$ at large disorder strengths due to finite background decays.}
	\label{SOMs_lifetimes_fig}
\end{figure}

Fig.~\ref{SOMs_lifetimes_fig} shows an analysis of the slow dynamics using simple exponentials of the form $\mathcal{I}(t) \sim e^{-\gamma t}$. The decay rates $\gamma$ are extracted via linear fits of $\log(\mathcal{I})$ versus $t$. The analysis is equivalent to the power-law analysis presented in Fig.~3 in the main text. We find that the qualitative results of increasingly slow dynamics with increasing disorder strengths, as well as the saturation to an offset value due to external decays do not change. Also, the quantitative estimation of the lower bound of the critical disorder strength to a value of $\Delta_c^\mathrm{MBL} \gtrsim 3.8 \pm 0.5 \, J$ is unchanged.
\\

\subsection{Analysis of background decay}

\begin{figure}
	\centering
	\includegraphics[width=84mm]{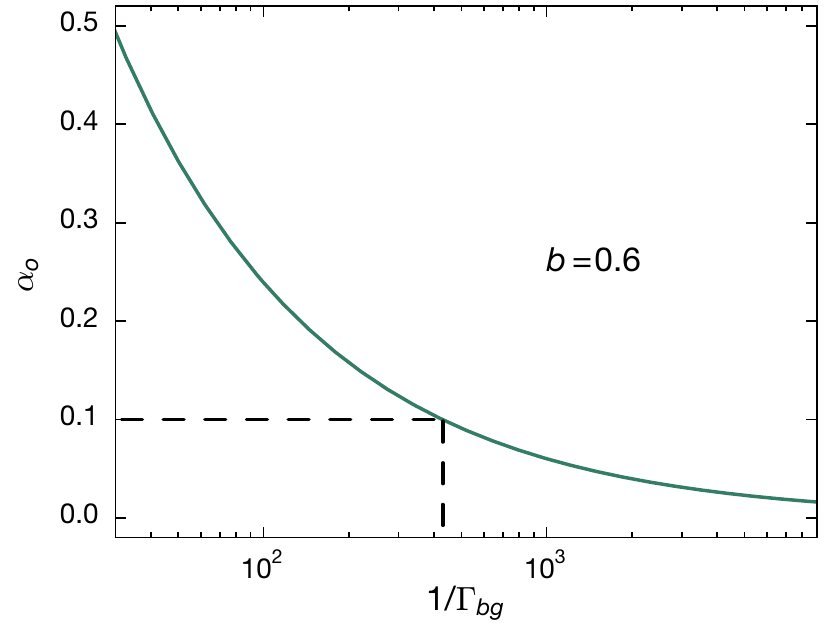}
	\caption{\textbf{Effect of background decay:} Exponents extracted by a power-law fit in the experimental time window of $8 - 40 \, \tau$ to a stretched exponential decay with rate $\Gamma_{bg}$ and stretching exponent $b=0.6$. The offset of $\alpha_o \approx 0.1$ we find in Fig.~3 corresponds to a background decay with a lifetime of $1/\Gamma_{bg} \approx 400-500 \, \tau$, as is indicated by the dashed lines.}
	\label{SOMs_background_fig}
\end{figure}

In the analysis of the power-law exponents (see Fig.~3) we find that at large $\Delta$, the exponents saturate at a nonzero offset value $\alpha_o$. This behavior is consistent with an underlying exponential long-term decay $\mathcal{I}(t) \sim e^{-(\Gamma_{bg} t)^b}$ of the imbalance. Here, $\Gamma_{bg}$ is the imbalance lifetime and $b$ the stretching exponent. Such stretched exponential decays have been found to arise in our system due to external bath couplings~\cite{Bordia16,Lueschen16}, such as photon scattering and residual tunnel couplings between the one dimensional tubes. Fig.~\ref{SOMs_background_fig} shows the result of fitting a power-law decay in the experimental time window of $8-40 \, \tau$ to a stretched exponential decay with $b = 0.6$ (a value we often found in previous works~\cite{Bordia16,Lueschen16}) and various $\Gamma_{bg}$. We find, that the observed offset of our exponents $\alpha_o$ corresponds to an imbalance lifetime of $1/\Gamma_{bg} \approx 400 - 500 \, \tau$, which is consistent with previous results~\cite{Bordia16,Lueschen16}.

\subsection{Usage of two different sets of data}

\begin{figure}
	\centering
	\includegraphics[width=84mm]{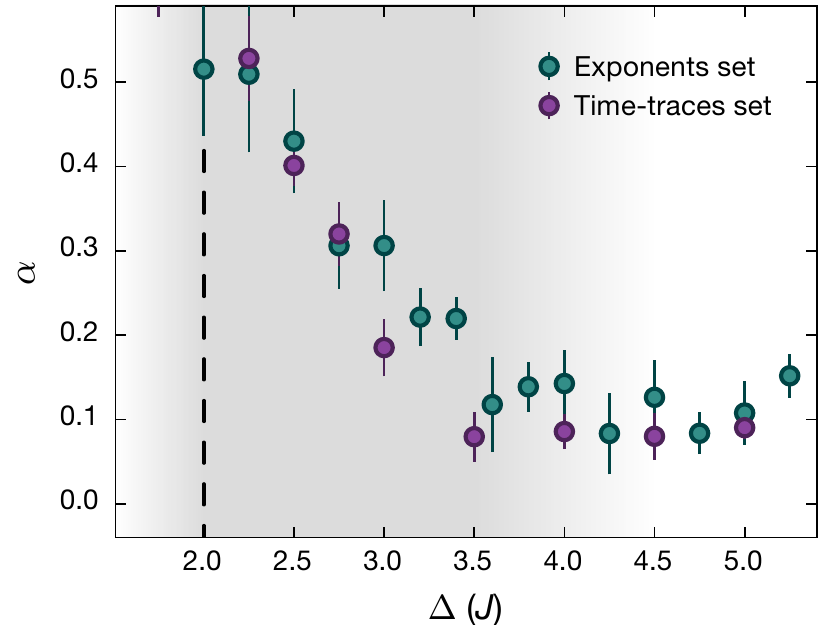}
	\caption{\textbf{Power law exponent extracted from time traces in Fig.~2:} Exponents extracted from the dataset used in Fig.~2 (Time-traces set) and in Fig~3 (Exponents set).}
	\label{SOMs_datasets_fig}
\end{figure}

In this work we use two different sets of data for the exponents in Fig.~3 and the time-traces in Fig.~2. The two sets were taken under the same experimental conditions, but approximately half a year apart and are compared in Fig.~\ref{SOMs_datasets_fig}. We find that the two sets agree within experimental noise.
The two datasets were employed as one of them contains more disorder strengths while the other uses finer time steps.

\subsection{Details of exact diagonalization simulations}

We follow the time evolution of the wave-function $| \Psi(t) \rangle = \exp(-i H \tau) | \Psi(0) \rangle$ from an initial CDW state  $| \Psi(0) \rangle$ in the Krylov space  ${\cal K}_m (H,| \Psi(0) \rangle)$, using a parallel solver as implemented in the SLEPc library~\cite{SLEPC}. The time evolution is essentially exact, with a convergence ensured by using small times steps, $d\tau$ in between $0.1$ and $1$ (we set $J=1$ in the simulations). We use chains of size $S=12, S=16$ and $S=20$, taking open boundary conditions and fixing $\beta = 0.721$. We work at zero magnetization and quarter-filling ($S/4$ up spin and $S/4$ down spin fermions) with Hilbert space sizes 48400 ($S=12$), 3312400 ($S=16$), and larger than 240 Millions ($S=20$). The initial state is a product state chosen of  CDW form, with a random repartition of up and down spins. We average results for the imbalance ${\cal I}(\tau)= \langle \Psi(\tau) | {\cal I} |  \Psi(\tau) \rangle$ over between $80$ and $200$ combinations of initial states and random values of the phase $\phi$ (chosen uniformly in $[0,2\pi]$). For the direct comparison to the experimental traces in Fig.~2, we scale the simulated imbalances by a factor of 0.9 to account for the independently measured initial imbalance in the experimental CDW state.

\begin{figure}
	\centering
	\includegraphics[width=84mm]{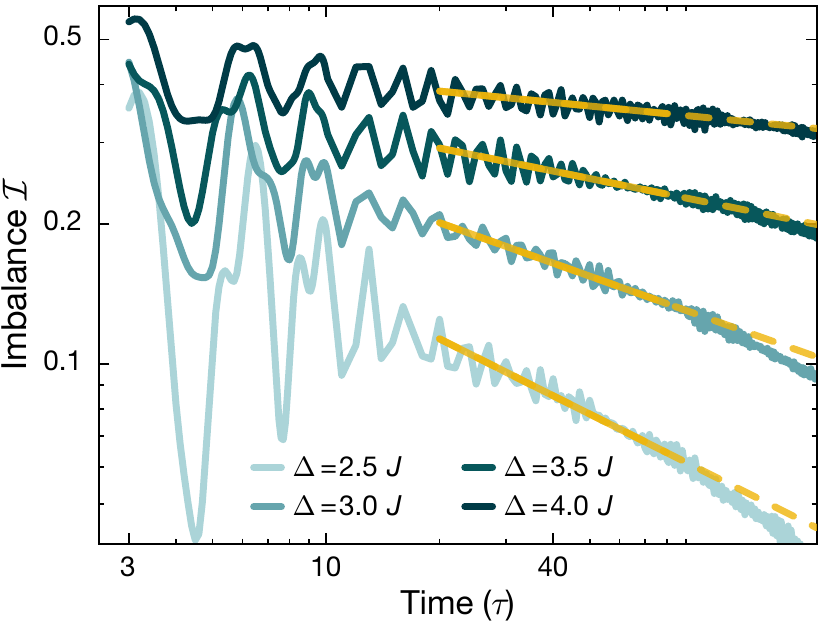}
	\caption{\textbf{ED time traces:} Decay of an initial charge-density wave as simulated on  systems with 20 sites and 10 atoms. Thickness of the lines denotes the statistical error after averaging over random initial conditions. The yellow lines illustrate fits to the data between $20-80\, \tau$, used to extract the exponents shown in Fig.~3. At longer times and for smaller values of disorder, a deviation from this power-law regime is observed.}
	\label{SOMs_EDtraces_fig}
\end{figure}

Fig.~\ref{SOMs_EDtraces_fig} shows exemplary traces, as well as the fitted power laws. The fits have been performed between $20-80 \, \tau$, as at shorter times the initial oscillations in the imbalance heavily affect the fit. In the experiment, this is not a problem, as the initial oscillations dephase quicker due to averaging over many 1D tubes with slightly different $J$~\cite{Schreiber15}. We observe that the power-law fits describe the behavior well until the imbalance decay becomes steeper (at around $100 \, \tau$ for the smaller disorder strengths). In the picture of state-induced rare regions this might be due to a redistribution of atoms and spins resulting in a "melting" of rare regions and therefore a faster decay at later times. Such a redistribution could occur by parts of the system thermalizing or due to a delocalized spin sector~\cite{Prelovsek16}. However, a recent study on quasiperiodic systems without initial state disorder has also observed power laws only on intermediate time scales~\cite{Lee17}. Understanding this regime will require additional work.

\begin{figure}
	\centering
	\includegraphics[width=84mm]{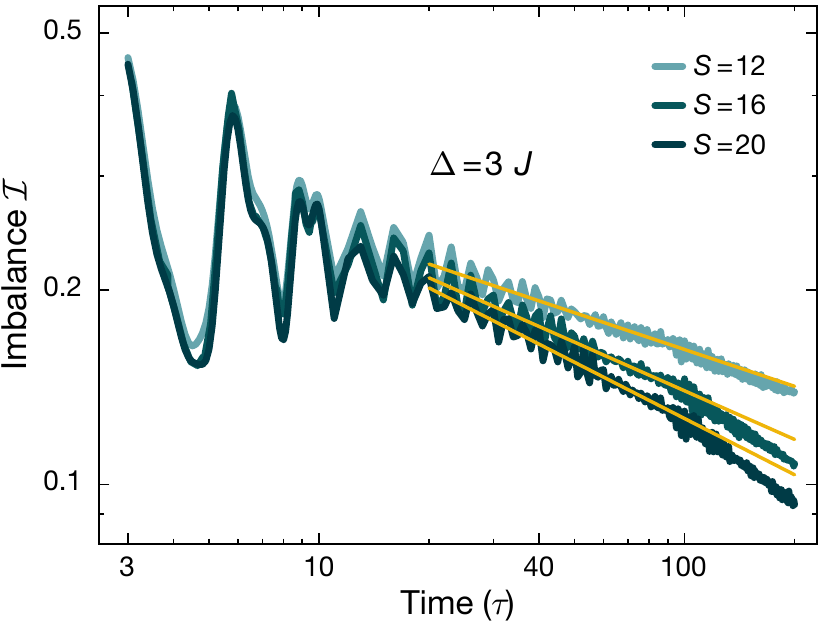}
	\caption{\textbf{Size dependence of ED traces:} Time evolution of the imbalance for various system sizes at $\Delta = 3 \, J$ and $U = 4\, J$. Thickness of the lines denotes the statistical error after averaging over random initial conditions. Yellow lines indicate power-law fits between $20-80 \, \tau$. We find a faster decay for larger system sizes.}
	\label{SOMs_finitesize_fig}
\end{figure}

Fig.~\ref{SOMs_finitesize_fig} shows time traces for $\Delta = 3 \, J$ and system sizes of $S=12,16,20$. We observe a slightly faster decay for larger system sizes. Due to the limited sizes available, we cannot extrapolate to infinite size. However, we find for all $S$ investigated, a systematic decrease of $1/z$ with increasing the disorder strength. We estimate the finite size error of the exponents indicated in Fig.~3 as the difference between the result of the $S=16$ and the $S=20$ sites simulation.
We find that this error decreases with increasing disorder strength. However, in the regime of large disorder ($\Delta \gtrsim 4J$) close to the transition, we cannot reach long enough times and large enough S to correctly estimate $\alpha$, and hence the systematic error of the exponent is much larger than the indicated finite size error, likely on the order of the exponent $\alpha \approx 0.05$ itself.

\end{document}